\documentclass{aa}
\usepackage{graphicx}
\usepackage{txfonts}
\usepackage{natbib}
\bibpunct{(}{)}{;}{a}{}{,}
\begin{document}
 
 \title{ATLASGAL-selected massive clumps in the inner Galaxy. IX. Deuteration of ammonia}

  \author{M. Wienen\inst{1}, F. Wyrowski\inst{1}, C. M. Walmsley\inst{2,}\inst{3}, T. Csengeri\inst{1}, T. Pillai\inst{1}, A. Giannetti\inst{4} \and K. M. Menten\inst{1}
               }

\institute{\inst{1}Max-Planck-Institut f\"ur Radioastronomie, Auf dem H\"ugel 69, 53121 Bonn, Germany\\ \email{mwienen@mpifr-bonn.mpg.de}\\
        \inst{2}Osservatorio Astrofisico di Arcetri, Largo E. Fermi, 5, I-50125 Firenze, Italy\\
        \inst{3}Dublin Institute of Advanced Studies, Fitzwilliam Place 31, Dublin 2, Ireland\\
   \inst{4}INAF - Istituto di Radioastronomia, Via Gobetti 101, 40129 Bologna, Italy}

   \date{Received }

  \abstract 
  % context heading (optional)
  {Deuteration has been used as a tracer of the evolutionary phases of low- and high-mass star formation. The APEX Telescope Large Area Survey (ATLASGAL) provides an important repository for a detailed statistical study of massive star-forming clumps in the inner Galactic disc at different evolutionary phases.}
  % aims heading (mandatory)
   {We study the amount of deuteration using NH$_2$D in a representative sample of high-mass clumps discovered by the ATLASGAL survey covering %from the earliest to the more evolved 
        various evolutionary phases of massive star formation. The deuterium fraction of NH$_3$ is derived from the NH$_2$D $1_{11}-1_{01}$ ortho transition at $\sim$ 86 GHz and NH$_2$D $1_{11}-1_{01}$ para line at $\sim$ 110 GHz.
        This is refined for the first time by measuring the NH$_2$D excitation temperature directly with the NH$_2$D $2_{12}-2_{02}$ para transition at $\sim$ 74 GHz. Any variation of NH$_3$ deuteration and ortho-to-para ratio with the evolutionary sequence is analysed.}
  % methods heading (mandatory)
   {Unbiased spectral line surveys at 3 mm were conducted towards ATLASGAL clumps between 85 and 93 GHz with the Mopra telescope and from 84 to 115 GHz using the IRAM 30m telescope. A subsample was followed up in the NH$_2$D transition at 74 GHz with the IRAM 30m telescope. We determined the deuterium fractionation from the column density ratio of NH$_2$D and NH$_3$ and measured the NH$_2$D excitation temperature for the first time from the simultaneous modelling of the 74 and 110 GHz line using MCWeeds. We searched for trends in NH$_3$ deuteration with the evolutionary sequence of massive star formation. We derived the column density ratio from the 86 and 110 GHz transitions as an estimate of the NH$_2$D ortho-to-para ratio.}
  % results heading (mandatory)
   {We find a large range of the NH$_2$D to NH$_3$ column density ratio up to $1.6 \pm 0.7$ indicating a high degree of NH$_3$ deuteration in a subsample of the clumps. Our analysis yields a clear difference between NH$_3$ and NH$_2$D rotational temperatures for a fraction. We therefore advocate observation of the NH$_2$D transitions at 74 and 110 GHz simultaneously to determine the NH$_2$D temperature directly. 
   	We determine a median ortho-to-para column density ratio of $3.7 \pm 1.2$.} 
   {The high detection rate of NH$_2$D confirms a high deuteration previously found in massive star-forming clumps. Using the excitation temperature of NH$_2$D instead of NH$_3$ is needed to avoid an overestimation of deuteration. We measure a higher detection rate of NH$_2$D in sources at early evolutionary stages. The deuterium fractionation shows no correlation with evolutionary tracers such as the NH$_3$ (1,1) line width, or rotational temperature. 
   }

    \keywords{Surveys --- Submillimeter --- Radio lines: ISM --- 
          ISM: molecules --- Stars: massive --- Stars: formation}
\titlerunning{ATLASGAL - NH$_3$ deuteration}
\authorrunning{M. Wienen et al.}
   \maketitle
%
%________________________________________________________________

\section{Introduction}
High-mass stars are known to form in dense clusters. They are much rarer than low-mass stars according to the stellar initial mass function \citep{2013pss5.book..115K} and are therefore located at greater distances. Massive protostars evolve embedded in dense cores \citep[$\sim 10^5-10^8$ cm$^{-3}$,][]{1999PASP..111.1049G,2000prpl.conf..299K} within high-mass star-forming complexes. These are more crowded than low-mass star-forming regions and have a short evolutionary timescale of $\sim 10^5$ yr \citep{2002Natur.416...59M}. These constraints restrict observations of the early phases of high-mass star formation. However, a key issue preventing a more complete understanding of the formation process of massive stars is the difficulty in revealing their initial conditions. 

The abundance of deuterium bound in molecules is orders of magnitude higher in cold molecular clouds than the primordial D/H ratio \citep[$\sim 10^{-5}$,][]{2003ApJ...587..235O}. Rising deuteration is expected from chemical models even into the gravitational collapse phase of the molecular cloud core \citep{2002P&SS...50.1133C,2017MNRAS.469.2602K}. Deuterated molecules can form through reactions between gaseous species as well as through the depletion of those onto grains with subsequent deuteration on the surfaces followed by the evaporation of icy grain mantles by the radiation from protostars back into the gas. At low temperatures ($<$ 20 K) and for low ortho-to-para H$_2$ ratios, deuterium fractionation is primarily regulated by reactive collisions 
\begin{eqnarray}
\label{H2D+reaction}
\rm H_3^+ + HD \rightarrow H_2D^+ + H_2 + \Delta E.
\end{eqnarray}
The production of H$_2$D$^+$ is essential for the deuterium chemistry, representing the first stage of deuterium enrichment \citep[``deuteration'',][]{2000A&A...361..388R,2005ChA&A..29..370W,2007AA...467..207P}. Gas-grain models comparing deuteration of H$_3^+$ at gas and dust temperatures of 10 K and 20 K by \cite{2015A&A...578A..55S} lead to a decrease in deuteration at the higher temperature, at which the reaction given in equation \ref{H2D+reaction} proceeds more efficiently in the backward direction. This trend is favoured by a high ortho-to-para H$_2$ ratio as well. Moreover, neutral molecules such as CO and H$_2$ destroy H$_2$D$^+$ at temperatures above 25 K and thus reduce the deuterium fraction \citep{2000A&A...361..388R}. According to the gas-grain models from \cite{2015A&A...578A..55S}, a total depletion of ammonia from the gas phase also occurs after $\sim 10^6$ years at a density of $10^6$ cm$^{-3}$ and a temperature of 20 K. In addition to ammonia, various forms of deuterated ammonia are also depleted onto grain surfaces with deuterium being trapped onto the surfaces \citep{2015A&A...581A.122S}. This leads to HD depletion and the decrease of the overall gas-phase deuteration efficiency \citep{2015A&A...578A..55S}.

It is known from observations and theory that C-bearing molecules such as CO freeze out onto dust grains in the cold and dense environment of molecular cores \citep{1999ApJ...523L.165C,1999A&A...342..257K,2002ApJ...569..815T,2005A&A...436..933F,2007ARA&A..45..339B,2014A&A...570A..65G} and therefore increase the [H$_2$D$^+$]/[H$_3^+$] ratio, which results in an enhanced abundance of deuterated species in the very early evolutionary phase. \cite{2014A&A...565A..75C} estimated that 25\% of the embedded sources in the ATLASGAL \citep{2009A&A...504..415S} sample with a peak intensity $> 5$ Jy are in the coldest stage of high-mass star formation. Such dense and cold clumps are therefore ideal targets to investigate their deuteration.

The amount of deuteration depends on the H$_2$ ortho-to-para ratio as well. If ortho instead of para H$_2$ is present, the backward reaction destroying H$_2$D$^+$ will be faster because of the four-times-larger ortho than para H$_2$ rate coefficients \citep{1992A&A...258..479P}. In addition, if the abundance of ortho H$_2$ is high, it will efficiently destroy H$_2$D$^+$ at low temperatures. A lower H$_2$ ortho-to-para ratio, as found in cold cores \citep{1992A&A...258..479P}, consequently leads to a higher deuterium enrichment.

NH$_2$D has been detected in low- and high-mass star-forming regions: it was observed in cold dark clouds by \cite{2000ApJ...535..227S}, in low-mass protostellar cores by \cite{2001ApJ...554..933S}, and in low-mass protostars by \cite{2003AA...403L..25H}. These authors measured NH$_3$ deuteration factors between 0.001 and 0.3 with similar errors of $\sim 25$\% on average, while interferometric observations with high angular resolution of NH$_2$D and NH$_3$ by \cite{2007AA...470..221C} found an enhanced deuterium fractionation of $0.5 \pm 0.2$ at high densities of 10$^6$ cm$^{-3}$ in the centre of a nearby Taurus core. In massive star-forming regions, NH$_2$D was observed for example in pre- and protocluster clumps by \cite{2007AA...467..207P} with half of the sample exhibiting a high deuterium fraction of $\geq 13$\%.

Analysis of the N$_2$D$^+$/N$_2$H$^+$ ratio in low-mass starless cores and protostars shows the predicted relation of a decreasing deuterium fractionation from the youngest objects immediately after the beginning of collapse to the more advanced evolutionary state of a Class 0 protostar \citep{2005ApJ...619..379C,2009A&A...493...89E}. It is suggested that the NH$_3$ deuteration increases in low-mass cores up to 20 K and is constant at higher temperatures \citep{2001ApJ...554..933S}. However, these authors only observed a small sample with large errors in the deuteration factors. Moreover, an enhanced N$_2$D$^+$/N$_2$H$^+$ ratio was also measured at the earliest evolutionary stages of high-mass star formation and a decline from high-mass starless core candidates to high-mass protostellar objects and ultracompact (UC) HII regions was found by \cite{2011A&A...529L...7F}. \cite{2010AA...517L...6B} were able to use the [NH$_2$D]/[NH$_3$] ratio as an evolutionary indicator in the environment of an ultracompact HII region (UCHIIR). \cite{2015A&A...575A..87F} compared NH$_3$ deuteration in high-mass cores with the evolutionary sequence and found that the [NH$_2$D]/[NH$_3$] ratio determined in massive starless cores, high-mass protostellar objects, and ultracompact HII regions does not decrease with the evolution of the cores. Two high-resolution studies \citep{2010AA...517L...6B,2011A&A...530A.118P} report that very close to high-mass protostars and UCHII regions (few 1000 AU), there is evidence of removal of deuterated NH$_3$. On large scales \cite{2015A&A...575A..87F} determined a deuterated fraction of NH$_3$ above 0.1 right up to the most evolved phase of their single-dish sample. This indicates no evidence of gas-phase removal on the envelope scales (several 10000 AU) up to the most advanced evolutionary stage. Existing single-dish and high-resolution data therefore suggest that deuteration is becoming inefficient (very little additional deuteration taking place) on large scales and appears to be more or less fully removed in the immediate vicinity of protostars and HII regions.

Previous studies of deuterated ammonia in high-mass star formation have only focused on small samples or one evolutionary stage. In the present paper, we determine the NH$_3$ deuteration of a representative sample of massive clumps that are detected by the ATLASGAL survey and observed in NH$_2$D as part of an unbiased spectral line survey. Our analysis focuses on the influence of temperature on deuteration. In particular, this sample covers various phases of high-mass star formation and allows us to analyse any dependence of the deuterium fractionation on evolutionary stage.

We present the NH$_2$D observations of the $\sim 86$ and $\sim 110$ GHz lines of ATLASGAL sources and the data reduction in Section \ref{data}. We derive the NH$_2$D column density, excitation, and NH$_3$ deuteration in Sect. \ref{results}. In addition, we measure the column density ratio of NH$_2$D at $\sim 86$ and $\sim 110$ GHz as an estimate of the ortho-to-para ratio. We compare the NH$_3$ and NH$_2$D temperatures and analyse any trend of NH$_3$ deuteration with evolutionary tracers in Sect. \ref{discussion}. Moreover, we study the dependence of the ortho-to-para ratio on the NH$_3$ deuteration, line width, and rotational temperature. Our NH$_2$D analysis is summarised in Sect. \ref{conclusions}.

\section{Observations}
\label{data}
The NH$_2$D data were observed within two unbiased spectral-line follow-up observations of large ATLASGAL subsamples. The first project covered 8 GHz centred on 89 GHz in the fourth quadrant with the Mopra telescope \citep{2019MNRAS.484.4444U} located near Coonabarabran in Australia at a latitude and longitude of -31.2678$^{\circ}$ and 149.0997$^{\circ}$. The second survey covered the whole 3 mm band in the first quadrant using the EMIR receiver at the IRAM 30m telescope\footnote{IRAM is supported by INSU/CNRS (France), MPG (Germany) and IGN (Spain)} \citep{2016A&A...586A.149C}. We summarise the NH$_2$D and NH$_3$ transitions, that are used in this article, with their spectroscopic properties in Table \ref{spectroscopic}.

\begin{table*}[htbp]
\begin{minipage}{\textwidth}
\caption[]{Properties of the NH$_2$D and NH$_3$ transitions.}
\label{spectroscopic}
\centering
\begin{tabular}{l c c c c c c l}
\hline\hline
Molecule & Quantum numbers & Frequency & upper energy level  & statistical weight & A$_{ij}$ & critical density & Reference \\ 
 & & (GHz) & (K) &  of upper/lower level & (1/s) & (1/cm$^{3}$) \\ \hline 
NH$_2$D ortho  &  1$_{11}$ - 1$_{01}$   &  85.926  &  20.7 & 27 & $7.82 \times 10^{-6}$ & $6.52 \times 10^4$\tablefootmark{a}  & CDMS \tablefootmark{c}\\
NH$_2$D para  & 1$_{11}$ - 1$_{01}$   &  110.154 & 21.3  & 9 & $1.65 \times 10^{-5}$ & $1.38 \times 10^5$\tablefootmark{a} & CDMS \\
NH$_2$D para & 2$_{12}$ - 2$_{02}$ & 74.156 & 50.7  & 15  & $5.92 \times 10^{-6}$ & $6.18 \times 10^4$\tablefootmark{a} & CDMS \\
NH$_3$ para  &  1,1 a - 1,1 s & 23.694   & 23.3  & 6  & $1.68  \times 10^{-7}$ & $1.95 \times 10^3$\tablefootmark{b}  & JPL \tablefootmark{c}\ \\ 
NH$_3$ para & 2,2 a - 2,2 s & 23.722 & 64.4 & 10 & $2.24  \times 10^{-7}$ & $2.03 \times 10^3$\tablefootmark{b}  & JPL \\ \hline 
\end{tabular}
\tablefoot{
        \tablefoottext{a}{The de-excitation rate coefficients are taken at 10 K from \cite{2014MNRAS.444.2544D}.}
        \tablefoottext{b}{The de-excitation rate coefficients at 15 K from \cite{1988MNRAS.235..229D} are used.}
   \tablefoottext{c}{The statistical weight given by CDMS includes a factor three for the rotation quantum number, a factor three for the spin of the two H atoms, and the N nuclear spin multiplicity of three. On the contrary, JPL does not include the nuclear spin of N.}}
\end{minipage}
\end{table*}

\subsection{Mopra observations}    
The Mopra 22m telescope was used to observe a 3mm molecular-line survey towards an unbiased ATLASGAL subsample of clumps with infrared association and peak fluxes above 1.75 Jy/beam at 870 $\mu$m as well as cold sources with peak fluxes above 1.2 Jy/beam \citep{2019MNRAS.484.4444U}. We observed 567 ATLASGAL sources located between $l = 300^{\circ}$ and 359$^{\circ}$ and $|b| \leq 1.5^{\circ}$ in 2008 and 2009.

This article focuses only on observations of the NH$_2$D 1$_{11}-1_{01}$ ortho transition at 85.926 GHz in the fourth quadrant. We used a 3mm HEMT receiver as frontend. Our measurements in the 3 mm band range from a frequency of $\sim 85.2$ GHz to $\sim 93.4$ GHz and were centered on 89.3 GHz. The UNSW Mopra spectrometer (MOPS) from the University of New South Wales contains four slightly overlapping 2.2 GHz bands leading to an overall $\sim 8$ GHz continuous bandwidth. We used MOPS in a broadband mode, where each 2.2 GHz wide band has a velocity resolution of 0.9 km~s$^{-1}$. The Mopra telescope has a beamwidth (FWHM) of 38$\arcsec$ at the frequency of the NH$_2$D line at $\sim 86$ GHz.

Pointed observations were conducted in position-switching mode. We examined the region around each source using ATLASGAL and infrared continuum maps from the Midcourse Space Experiment \citep[MSX,][]{2001AJ....121.2819P} and chose an offset position that is free of continuum emission at 20 $\mu$m, either $\pm 5\arcmin$ in longitude or latitude. We observed two polarisations of the NH$_2$D line at 86 GHz simultaneously. The total integration time for each source was $\sim 15$ min, resulting in an rms noise level of 24 mK on average at a velocity resolution of 0.9 km~s$^{-1}$. The median system temperature was about 200 K. Pointing was measured each hour with line pointings on SiO masers and a reference spectrum of G327 and M17 was obtained each day.

We processed the data initially with the ASAP package, which consisted of processing of the on-off observing mode, the time and polarisation averaging, and baseline subtraction. We converted the data to the $T_{\rm A}^*$ temperature scale and exported the data to the CLASS software from the GILDAS package\footnote{available at http://www.iram.fr/IRAMFR/GILDAS} for subsequent analysis. For the calibration from $T_{\rm A}^{*}$ to $T_{\rm MB}$ we corrected for the beam efficiency of 0.49 \citep{2005PASA...22...62L}.

\begin{figure*}
\centering
\includegraphics[angle=0,width=13.0cm]{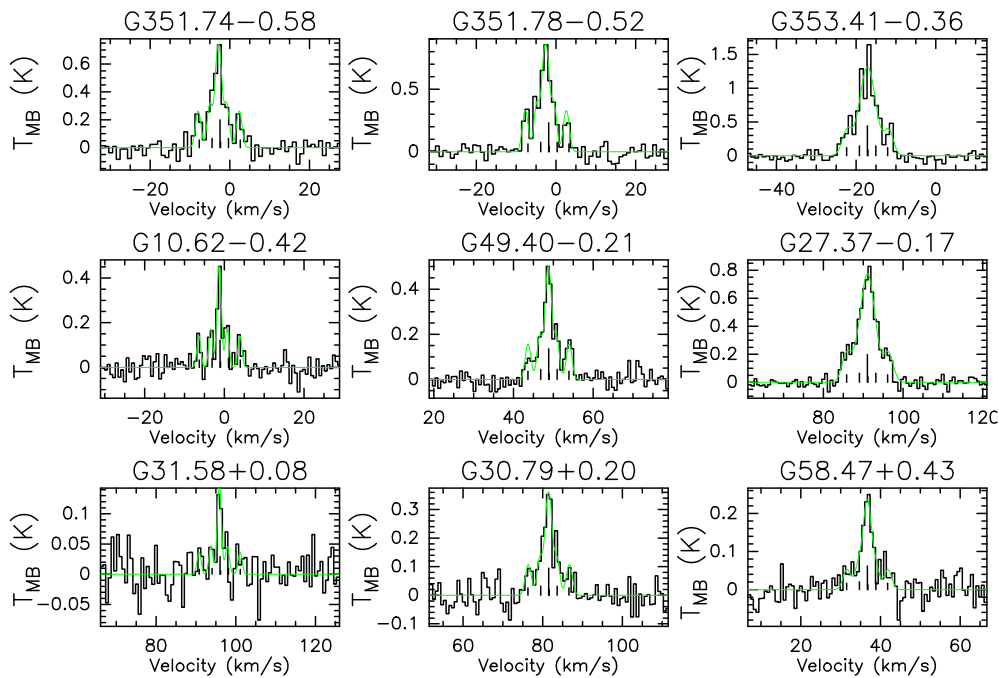}
\caption[Spectra of observed NH$_2$D ortho lines]{Examples of reduced and calibrated spectra of observed NH$_2$D transitions at 86 GHz; the fit is shown in green. The hyperfine structure of the spectra in the first and second rows are clearly detected, while that of the spectra in the third row is too weak to be visible. Frequencies of the hyperfine structure components are indicated by straight lines.}\label{nh2dlines-ortho}
\end{figure*}

\begin{figure*}
\centering
\includegraphics[angle=0,width=13.0cm]{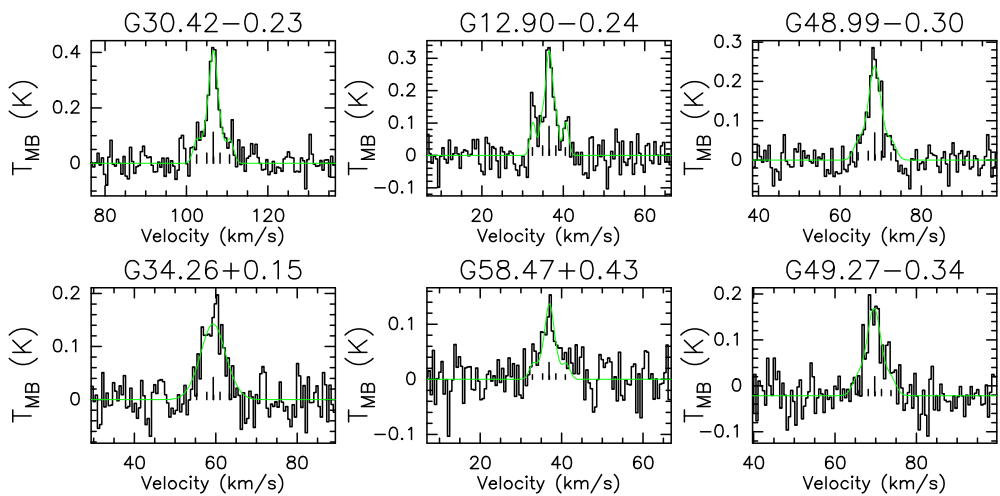}
\caption[Spectra of observed NH$_2$D para lines]{Examples of reduced and calibrated spectra of observed NH$_2$D lines at 110 GHz; the fit is indicated in green. Frequencies of the hyperfine structure components are labelled.}\label{nh2dlines-para}
\end{figure*}

\begin{figure}[h!]
\centering
\includegraphics[angle=0,width=9.0cm]{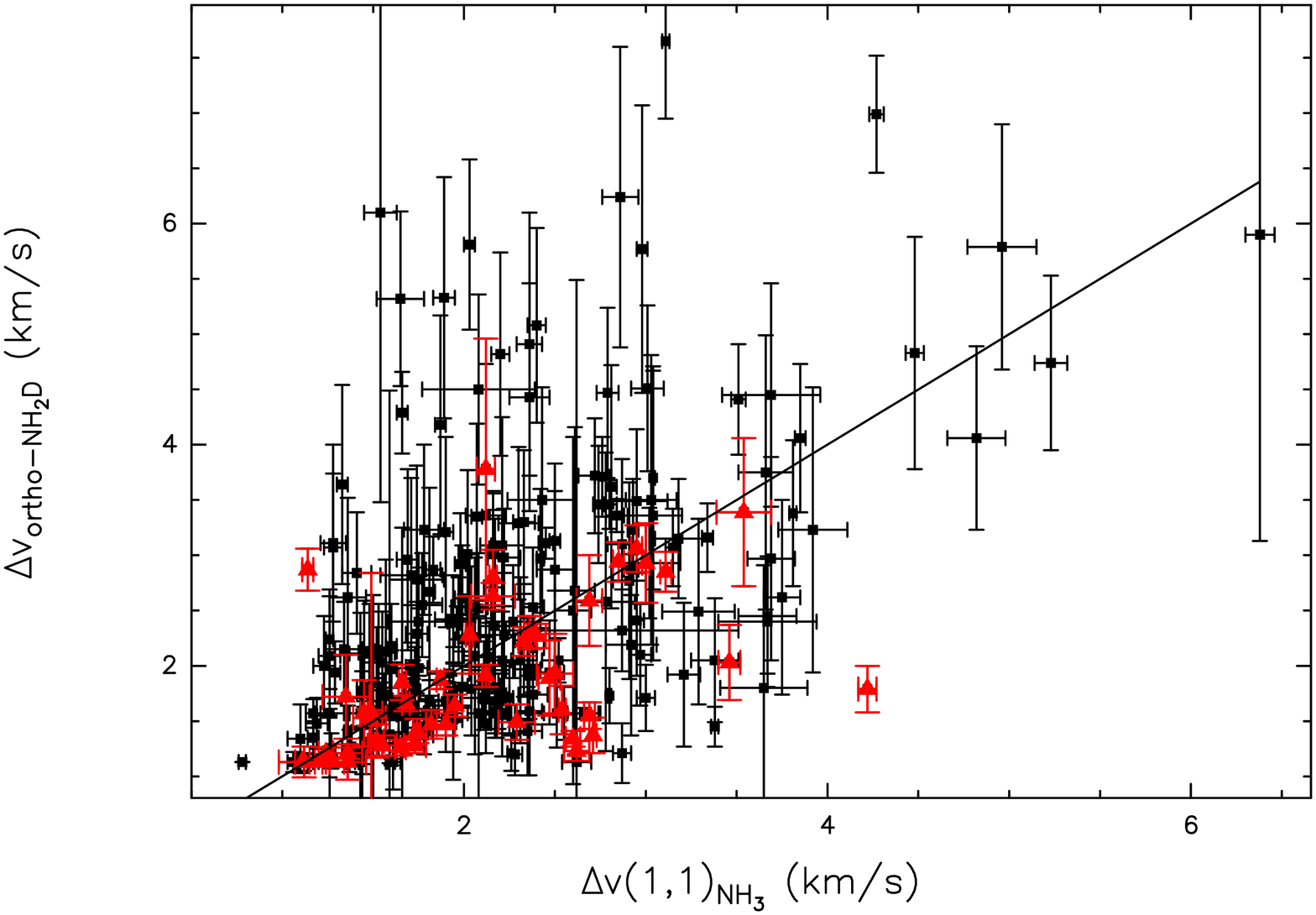}
\caption[Ortho NH$_2$D linewidth plotted against NH$_3$ (1,1) linewidth]{Line width of the NH$_2$D transition at 86 GHz plotted against the NH$_3$ (1,1) line width. ATLASGAL sources that have an error in the NH$_2$D optical depth of less than or greater than 50\% are indicated as red triangles or black points, respectively. The straight line corresponds to equal line widths.}\label{dvnh2d-dv11}
\end{figure}
    
\subsection{IRAM 30m observations}    
The NH$_2$D 1$_{11}-1_{01}$ ortho line at 85.926 GHz and the para transition at 110.154 GHz in the first quadrant were measured as part of the large molecular line survey of ATLASGAL sources conducted with the IRAM 30m telescope in 2011 and 2012 \citep{2016A&A...586A.149C}. The sample covers a range of evolutionary phases from the quiescent clumps to actively star-forming clumps hosting HII regions. The IRAM sample targeted bright sources, but also includes infrared-quiet clumps meaning without a detection at 22 $\mu$m corresponding to roughly half of the targeted sources \cite[see more details in][]{2016A&A...586A.149C}. While the Mopra sample consists of clumps with and without infrared association based on MSX data at 21 micron, the IRAM sample used the WISE 22 micron point-source catalogue with higher resolution. In a pilot study, 36 sources with the highest submillimetre (submm) peak flux densities from the ATLASGAL survey were observed in April 2011 and a second large sample was followed up in February, March, and October 2012. Using the IRAM 30m telescope we observed 425 ATLASGAL sources in NH$_2$D within $l = 5^{\circ} - 60^{\circ}$ and $|b| \leq 1.5^{\circ}$.

The observations were carried out with the EMIR receiver E090. The frequency range from $\sim 84$ to 115 GHz was divided into 4 GHz blocks for the pilot study, while the sample in 2012 was observed with a total bandwidth of 16 GHz and in two setups centred on 88 and 96 GHz. The Fast Fourier Transform Spectrometer (FFTS) was used with a spectral resolution of 200 kHz resulting in a velocity resolution of 0.68 km~s$^{-1}$ at $\sim 86$ GHz and of 0.53 km~s$^{-1}$ at $\sim 110$ GHz. The half-power beam width at the NH$_2$D 1$_{{11}-1_{01}}$ line frequencies at $\sim 86$ GHz and $\sim 110$ GHz is 29$\arcsec$ and 22$\arcsec$. The spectra were converted to the  main beam brightness temperature scale for the beam efficiency of 0.81 at $\sim 86$ and 110 GHz as in \cite{2016A&A...586A.149C}.

The observations were conducted in position switching mode with a constant offset of 10$\arcmin$ in right ascension and declination with a total integration time of $\sim 4.5$ min for each source. Pointing and focus were measured regularly. G34.26+0.15 was mostly used as a spectral line calibrator.

The NH$_2$D $2_{12}-2_{02}$ para transition at 74.156 GHz was observed toward a subsample of the 24 brightest clumps in deuterated ammonia that was selected from the NH$_2$D observations at $\sim 86$ GHz covering different evolutionary phases such as 24$\mu$m dark sources, active clumps in IRDCs, and HII regions. The EMIR receiver with a frequency range between 71 and 79 GHz was used for position-switching observations toward the peaks of the clumps. Typical system temperatures were about 180 K. We measure an rms noise level of 15 mK at a velocity resolution of 0.75 km~s$^{-1}$, which is similar to the observations at $\sim 86$ GHz. A total integration time of $\sim 60$ minutes was spent per source including on and off position.

\begin{table*}
\begin{minipage}{\textwidth}
\caption[Parameters of the ortho NH$_2$D line]{Parameters of the NH$_2$D line at 86 GHz and NH$_3$ rotational temperature \citep{2018A&A...609A.125W}. Errors are given in parentheses. The full table is available at CDS.}              % title of Table
\label{parline-nh2d85}    
\centering                                     
\begin{tabular}{l c c l c c c c}          
\hline\hline                        
 & RA\tablefootmark{d} & Dec\tablefootmark{d} & $\tau$(1,1)\tablefootmark{e} & v$_{\rm LSR}$ & $\Delta \rm v$ & $T_{\mbox{\tiny MB}}$ & $T_{\mbox{\tiny rot}}$ \\ 
Name  & (J2000) &  (J2000) &  & (km~s$^{-1}$) &  (km~s$^{-1}$) & (K) & (K) \\                    
\hline
%Added by TeX Support                                 
G10.62-0.42 & 18 10 36.92 &  -19 57 00.86 & 0.94 $(\pm$0.36) & -1.28 $(\pm$0.04) & 1.13 $(\pm$0.03) & 0.43 $(\pm$0.04) & 15.7 $(\pm$0.8) \\
G27.37-0.17 & 18 41 51.25 & -05 01 42.71 & 0.72 $(\pm$0.18) & 91.05 $(\pm$0.05) & 2.85 $(\pm$0.18) &  0.77 $(\pm$0.04) & 19.5  $(\pm$0.7) \\
G30.42-0.23 & 18 47 40.32 & -02 20 29.49 & 0.93 $(\pm$0.19) & 105.61 $(\pm$0.01)  & 2.22 $(\pm$0.13) & 0.74 $(\pm$0.04) & 18.7 $(\pm$0.8) \\
G30.79+0.20 & 18 46 47.69 & -01 49 01.99 & 0.73 $(\pm$0.43)* & 81.51 $(\pm$0.10) &  1.76 $(\pm$0.27) & 0.36 $(\pm$0.04) & 18.5 $(\pm$0.8) \\
G31.58+0.08 & 18 48 41.95 & -01 10 00.53 & 0.38 $(\pm$1.02)* & 95.91 $(\pm$0.13) & 1.21 $(\pm$0.27) & 0.13 $(\pm$0.04) & 23.3 $(\pm$1.1) \\
G34.26+0.15 & 18 53 18.60 & +01 14 57.91 & 0.12 $(\pm$0.09)* &57.52 $(\pm$0.08) & 5.6 $(\pm$0.32) & 0.28 $(\pm$0.01) & 24.9 $(\pm$1.9) \\
G48.99-0.30 & 19 22 26.19 & +14 06 37.09 & 0.21 $(\pm$0.19)* & 67.92 $(\pm$0.07) & 2.87 $(\pm$0.23) & 0.50 $(\pm$0.02) & 23.0 $(\pm$0.9) \\
G49.27-0.34 & 19 23 07.06 & +14 20 14.86 & 0.15 $(\pm$0.64)* &68.44 $(\pm$0.12) & 3.23 $(\pm$0.46) & 0.73 $(\pm$0.06) & 20.7 $(\pm$1.0) \\
G49.40-0.21 & 19 22 55.75 & +14 30 49.66 & 0.80 $(\pm$0.25) & 48.78 $(\pm$0.10) & 1.64 $(\pm$0.10) & 0.49 $(\pm$0.04) &  14.4 $(\pm$0.7) \\
G58.47+0.43 & 19 38 57.86 & +22 46 37.76 & 0.10 $(\pm$0.27)* & 36.66  $(\pm$0.13) & 2.23 $(\pm$0.24) & 0.22 $(\pm$0.04) & 20.4 $(\pm$1.1) \\
G305.23-0.02 & 13 11 36.41 & -62 48 19.4 & 0.58 $(\pm$0.77)* & -29.48 $(\pm$0.15) & 1.57 $(\pm$0.31) & 0.27 $(\pm$0.06) & 17.4 $(\pm$2.3) \\
G305.82-0.11 & 13 16 47.85 & -62 50 36.3 & 0.10 $(\pm$0.88)* & -41.87 $(\pm$0.22) & 2.12 $(\pm$0.44) & 0.24 $(\pm$0.04) & 13.4 $(\pm$1.8) \\
G309.38-0.13 & 13 47 22.85 & -62 18 06.7 & 0.10 $(\pm$0.26)* & -51.14 $(\pm$0.47) & 2.50 $(\pm$1.57) & 0.22 $(\pm$0.04) & 16.1 $(\pm$1.5) \\
G310.01+0.39 & 13 51 38.30 & -61 39 14.3 & 2.86 $(\pm$1.82)* & -41.18 $(\pm$0.26) & 1.57 $(\pm$0.30) & 0.18 $(\pm$0.06) & 18.6 $(\pm$1.7) \\
G351.74-0.58 & 17 26 47.69 & -36 12 07.56 & 1.18 $(\pm$0.28) & -02.81 $(\pm$0.06) & 1.57 $(\pm$0.30) & 0.73 $(\pm$0.04) & 15.2 $(\pm$0.7) \\
G351.78-0.52 & 17 26 39.16 & -36 08 04.58 & 1.43 $(\pm$0.28) & -02.61 $(\pm$0.05) & 1.66 $(\pm$0.07) & 0.86 $(\pm$0.04) & 15.1 $(\pm$0.5) \\
G353.41-0.36 & 17 30 26.87 & -34 41 50.53 & 0.88$(\pm$0.20) & -17.13 $(\pm$0.07) & 2.87 $(\pm$0.19) & 1.31 $(\pm$0.04) & 14.7 $(\pm$1.3) \\
\hline                                           
\end{tabular}
\tablefoot{
        \tablefoottext{d}{Units of right ascension are hours, minutes, and seconds, and units of declination are degrees, arcminutes, and arcseconds.}
    \tablefoottext{e}{The smallest optical depth given by the CLASS software is 0.1. Sources without detected hyperfine structure or no reliable derivation of the optical depth due to low S/N (see Sect. \ref{deuteration calculation}) are marked by a star.}}
\end{minipage}
\end{table*}

\subsection{Data reduction}
The CLASS software was used to reduce the NH$_2$D data. To remove the baseline from the spectra we subtracted a polynomial baseline of order zero from the spectra, excluding velocity windows that were placed around the NH$_2$D lines. The hyperfine structure of the NH$_2$D transitions at 86 and 110 GHz was fitted taking six hyperfine components into account. The fit of the line at 110 GHz kept the line width as a fixed parameter using the NH$_2$D line width at 86 GHz assuming that the two transitions originate from the same gas. This gives the optical depth of the main line, $\tau$, the radial velocity, $\rm v_{\rm LSR}$, and the line width, $\Delta \rm v$, at the full width at half maximum of a Gaussian profile with their errors as the formal fit errors from CLASS. As the line width of the transition at 110 GHz is set to a fixed value, no error is indicated for this parameter in Table \ref{parline-nh2d110}. The temperature of the NH$_2$D line was measured from the peak of the hyperfine structure fit. The minimum optical depth of the hyperfine structure fit in CLASS is 0.1. Because the hyperfine structure of some NH$_2$D lines at 86 GHz and of most transitions at 110 GHz are too weak to be detected, we cannot determine their optical depth. In these cases, the fit from the CLASS software gives an error of the optical depth of greater than 50\%. The 86 GHz-NH$_2$D line parameters with the NH$_3$ rotational temperature between the (1,1) and (2,2) inversion transition from \cite{2012A&A...544A.146W} and \cite{2018A&A...609A.125W} are given in Table \ref{parline-nh2d85}. We report the molecular line parameters of the para NH$_2$D transition in Table \ref{parline-nh2d110} which lists the line-of-sight velocity of para NH$_2$D, $\rm v_{LSR_{110}}$, the line width, $\Delta \rm v_{110}$, and the main beam brightness temperature, $T_{\rm MB_{110}}$.

\begin{table*}
        \begin{minipage}{\textwidth}
                \caption[Parameters of the ortho NH$_2$D line]{Line parameters of the NH$_2$D transition at 110 GHz with errors noted in parentheses. We made the full table available at CDS.}
                % title of Table
                \label{parline-nh2d110}      
                \centering                                      
                \begin{tabular}{l c c c c c}          
                        \hline\hline                        
                        & RA\tablefootmark{f} & Dec\tablefootmark{f} &  $\rm v_{LSR_{110}}$ & $\Delta \rm v_{110}$ &  $T_{\rm MB_{110}}$ \\ 
                        Name  & (J2000) &  (J2000) &   (km~s$^{-1}$) &  (km~s$^{-1}$) &  (K) \\                    
                        \hline 
                        %Added by TeX Support                                
                        G14.33-0.64 & 18 18 54.59 & -16 47 41.16 & 22.72$(\pm$0.09) & 3.06 & 0.49$(\pm$0.05) \\
G19.88-0.54 & 18 29 14.53 & -11 50 25.67 & 43.67$(\pm$0.31) & 2.27 & 0.33$(\pm$0.07) \\
G27.37-0.17 & 18 41 51.25 & -05 01 42.71 & 90.82$(\pm$0.15) & 2.85 & 0.38$(\pm$0.05) \\
G30.42-0.23 & 18 47 40.32 & -02 20 29.49 & 106.59$(\pm$0.08) & 2.22 & 0.36$(\pm$0.04) \\
G31.41+0.31 & 18 47 34.23 & -01 12 44.67 & 97.15$(\pm$0.311) & 5.12 & 0.33$(\pm$0.06) \\
G34.26+0.15 & 18 53 18.53 & 01 14 57.9 & 59.46$(\pm$0.20) & 5.52 & 0.18$(\pm$0.02) \\
G48.99-0.30 & 19 22 26.19 & +14 06 37.09 & 68.56$(\pm$0.12) & 2.87 & 0.27$(\pm$0.04) \\
G49.27-0.34 & 19 23 07.06 & +14 20 14.85 & 69.60$(\pm$0.22) & 3.23 & 0.25$(\pm$0.05) \\
G58.47+0.43 & 19 38 57.86 & +22 46 37.76 & 37.0$(\pm$0.27) & 2.23 & 0.11$(\pm$0.02) \\
                        \hline                                             
                \end{tabular}
                \tablefoot{
                        \tablefoottext{f}{Units of right ascension are hours, minutes, and seconds, and units of declination are degrees, arcminutes, and arcseconds.}
                        }
        \end{minipage}
\end{table*}

\begin{table*}
        \begin{minipage}{\textwidth}
                \caption[Parameters of the ortho NH$_2$D line]{Parameters of the NH$_2$D line at 74 GHz. Errors are given in parentheses. The full table is available at CDS.}
                % title of Table
                \label{parline-nh2d74}     
                \centering                                      
                \begin{tabular}{l c c c c c}          
                        \hline\hline                        
                        & RA\tablefootmark{g} & Dec\tablefootmark{g} &   $\rm v_{LSR_{74}}$ & $\Delta \rm v_{74}$ &  $T_{\rm rot_{NH_2D}}$ \\ 
                        Name  &  (J2000) &  (J2000) &   (km~s$^{-1}$) &  (km~s$^{-1}$) &  (K) \\                   
                        \hline 
                        %Added by TeX Support                                 
                        G12.50-0.22 & 18 13 41.49 & -18 12 35.51 & 36.31 & - & 17.7 \\
G14.33-0.64 & 18 18 54.59 & -16 47 41.16 & 22.75($^{+0.20}_{-0.19})$ & 2.7($^{+0.5}_{-0.6}$) & 22.8 ($^{+4.9}_{-5.5}$) \\
G19.88-0.54 & 18 29 14.53 & -11 50 25.67 & 41.87($^{+0.57}_{-0.6})$  & 4.1 ($^{+1.4}_{-1.4}$) & 20.0 ($^{+7.0}_{-5.0}$) \\
G23.21-0.38 & 18 34 55.02 & -08 49 16.96 & 77.87 & - & 58.9 \\
G27.37-0.17 & 18 41 51.25 & -05 01 42.71 & 90.76($^{+0.28}_{-0.29}$) & 3.5 ($^{+1.0}_{-1.0}$) & 18.7 ($^{+4.9}_{-6.0}$) \\
G30.85-0.08 & 18 47 55.54 & -01 53 33.38 & 98.27 & - & 12.0 \\
G31.41+0.31 & 18 47 34.23 & -01 12 44.67 & 96.05($^{+0.54}_{-0.48})$  & 5.9  ($^{+1.1}_{-0.9}$) & 36.5  ($^{+13.4}_{-10.3}$) \\
G34.26+0.15 & 18 53 18.53 & 01 14 57.9 & 59.11($^{+0.66}_{-0.74}$)  & 5.7 ($^{+1.8}_{-1.5}$) & 41.6 ($^{+13.8}_{-6.6}$) \\

                        \hline                                             
                \end{tabular}
                \tablefoot{
                        \tablefoottext{g}{Units of right ascension are hours, minutes, and seconds, and units of declination are degrees, arcminutes, and arcseconds.}
                }
        \end{minipage}
\end{table*}

\section{Results and analysis of the NH$_2$D sample}
\label{results}
\subsection{NH$_2$D detection rates}
We observed 992 ATLASGAL clumps in NH$_2$D 1$_{11}$ - 1$_{01}$  at 86 GHz in the first and fourth quadrant and detected NH$_2$D in 390 clumps (39\%) corresponding to a S/N ratio $>$ 3. We found that the NH$_2$D velocities of all detected sources are within $\sim 2.5$ km~s$^{-1}$ of the NH$_3$ velocities. 

The hyperfine structure of the NH$_2$D line at 86 GHz is clearly visible in 79 clumps in the first and fourth quadrants (20\%). We can determine the optical depth of these sources, which ranges between $0.34 \pm 0.02$ for G15.22$-$0.43 and $4.2 \pm 2.0$ for G35.58+0.01 with a median optical depth of the NH$_2$D line at 86 GHz of $1.12 \pm 0.35$. A few spectra of the NH$_2$D transition at 86 GHz of these sources and of the clumps, for which the hyperfine structure is too weak to obtain an optical depth, are shown in Fig. \ref{nh2dlines-ortho}.

The NH$_2$D 1$_{11}$ - 1$_{01}$ line at 110 GHz was observed in 373 ATLASGAL sources in the first quadrant of which 65 clumps are detected (17\%) with S/N  $>$ 3 and the NH$_2$D and NH$_3$ velocities of all detections are within $\sim 2.5$ km~s$^{-1}$. The hyperfine structure components of most clumps are blended and we detect those in only seven sources (2\%) with an optical depth between $0.12 \pm 0.02$ for G27.37$-$0.17 and $3.1 \pm 1.3$ for G23.44$-$0.18. The median optical depth of these clumps that show hyperfine structure is $1.1  \pm 0.8$. Examples of the NH$_2$D transition at 110 GHz are presented in Fig. \ref{nh2dlines-para}.

\begin{figure*}
\centering
\includegraphics[angle=0,width=9cm]{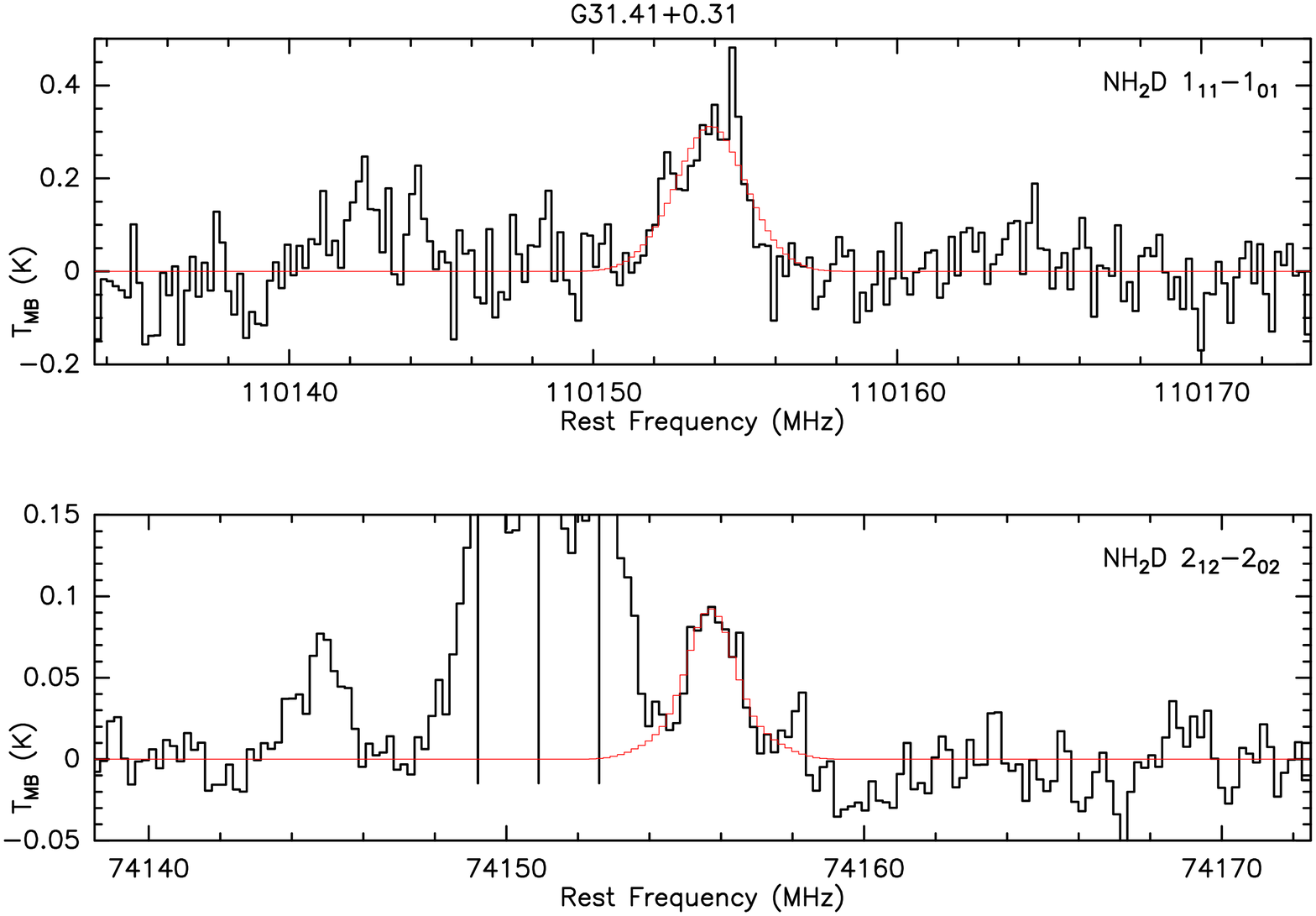}\hspace*{1cm}
\includegraphics[angle=0,width=9.0cm]{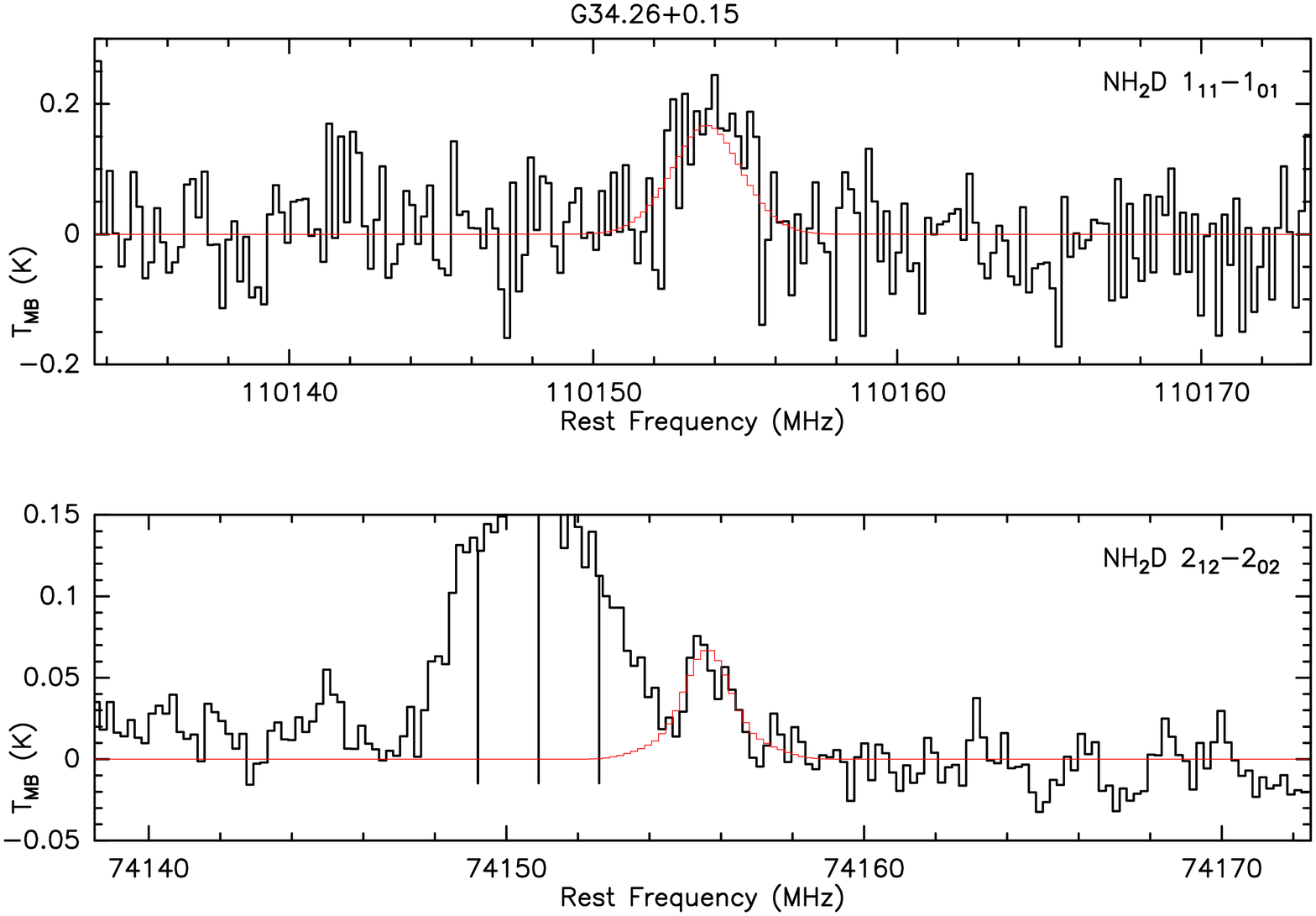}\vspace*{0.5cm}
\includegraphics[angle=0,width=9.0cm]{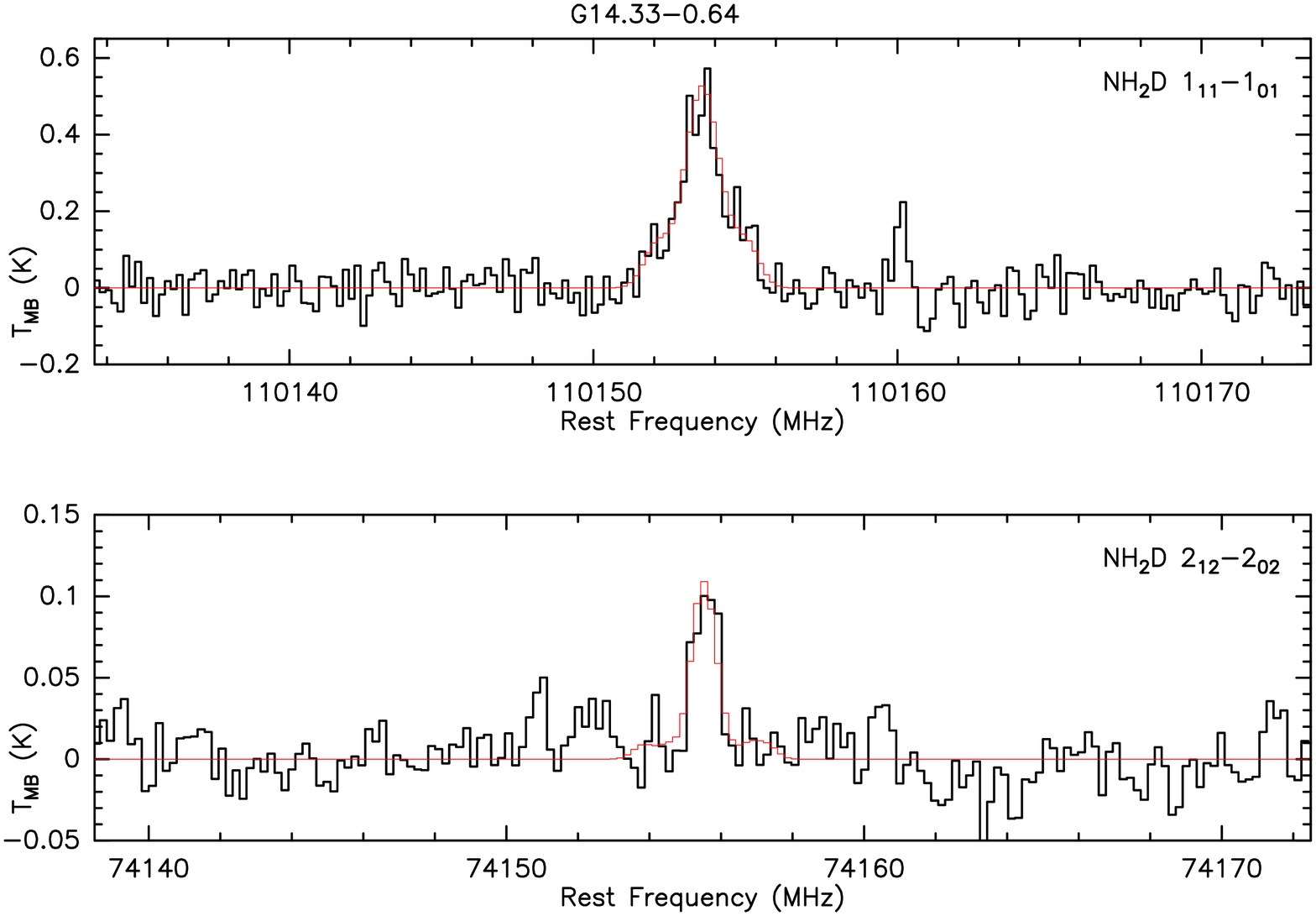}\hspace*{1cm}
\includegraphics[angle=0,width=9.0cm]{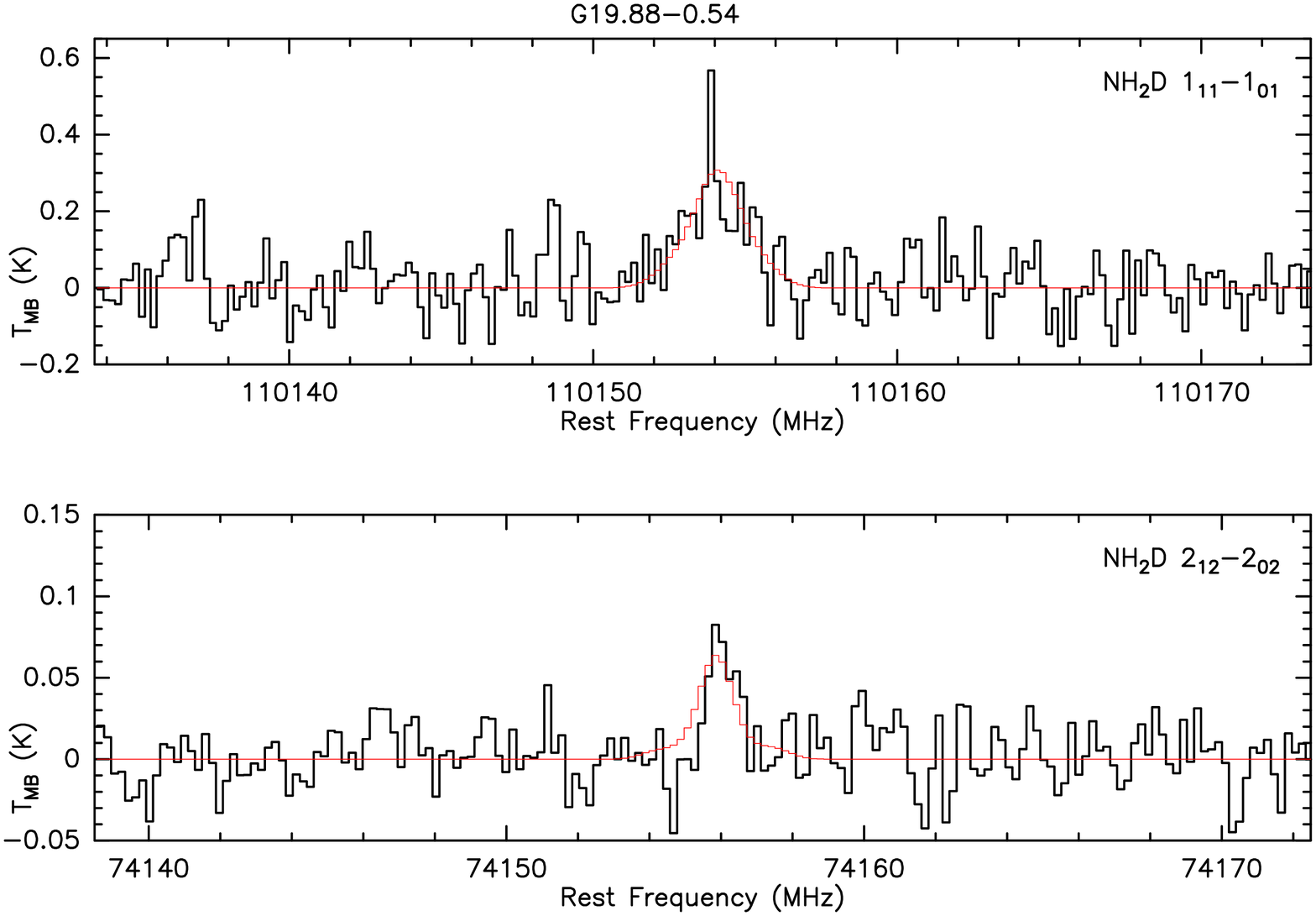}\vspace*{0.5cm}
\includegraphics[angle=0,width=9.0cm]{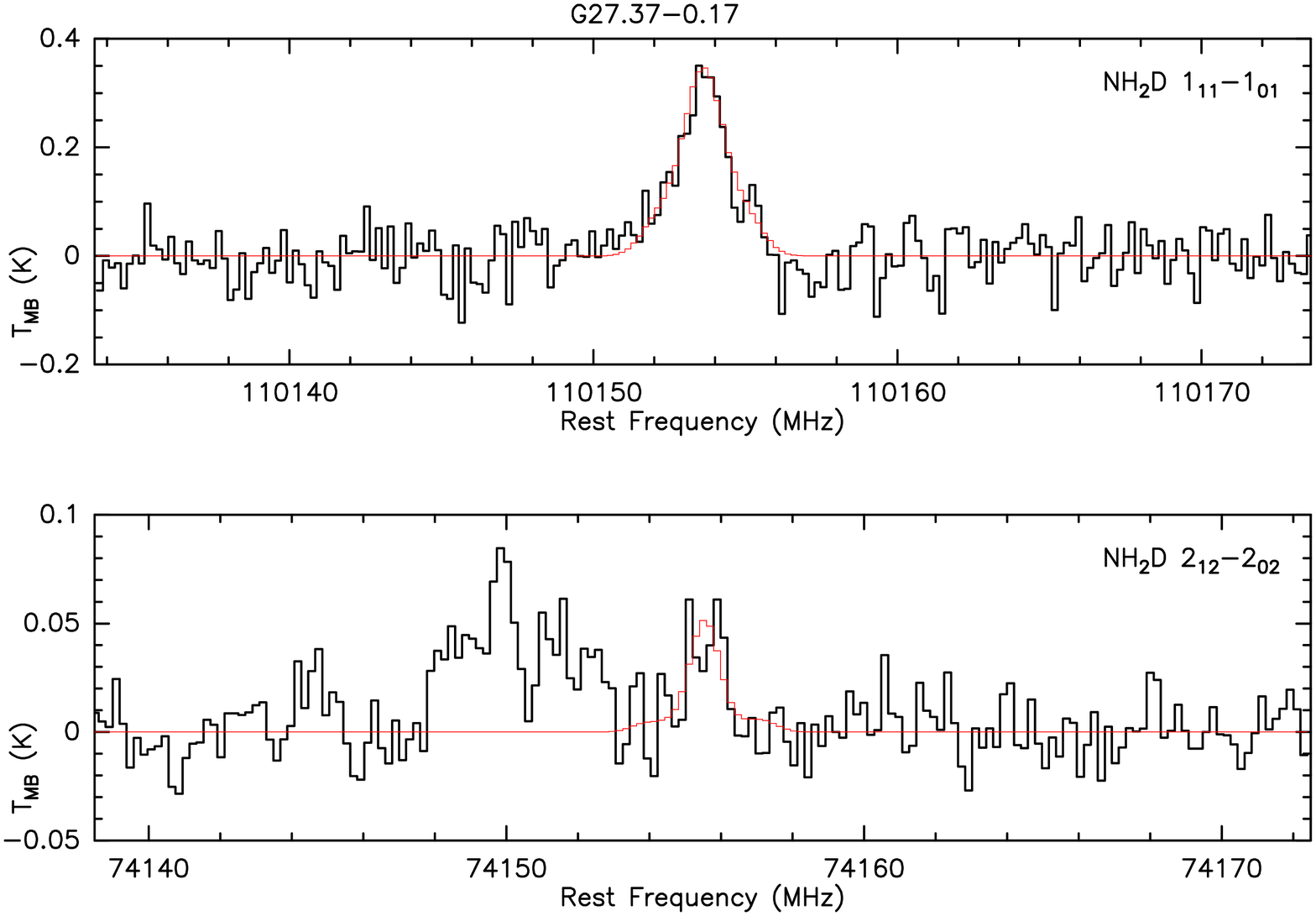}
\caption[Spectra and models of NH$_2$D lines]{Examples of reduced and calibrated spectra of observed NH$_2$D lines at 110 GHz and 74 GHz. The bright line of G31.41+0.31 and G34.26+0.15 with a frequency close to 74.15 GHz consists of four blended lines of CH$_3$OCH$_3$; their frequencies are labelled as lines in the spectra. Results of simultaneous modelling of the NH$_2$D transitions at 110 and 74 GHz (see Sect. \ref{nh2d excitation}) are illustrated in red.}\label{nh2d-lines-model}
\end{figure*}

We measured the NH$_2$D transition at 74 GHz in 24 ATLASGAL sources in the first quadrant and 5 sources have S/N $>$ 3. These detections possess NH$_2$D and NH$_3$ velocities that again lie close together, within $\sim 2.5$ km~s$^{-1}$. The results of the modelling of five detections are plotted in Fig. \ref{nh2d-lines-model}.

\subsection{NH$_2$D line width}\label{sec:nh2d-linewidth}
We derive the NH$_2$D intrinsic line width from hyperfine structure fits to the transition at 86 GHz, and find values between 1.1 km~s$^{-1}$ and 7.7 km~s$^{-1}$. Fitting of NH$_2$D lines with a small S/N and an error in the optical depth of greater than 50\% gives optical depths varying between 0.1 and 2.5, which causes the line width to vary by $\sim 41$\%, leading to a systematic error of the same order as the median error in the line width. In those cases and for an even larger error in the optical depth the fit would result in a large uncertainty on the line width. For sources with a smaller error in the optical depth, that is $< 50 $\%, the line width derived from the fit is reliable. We searched for a counterpart to the NH$_2$D observations within the NH$_3$ sample of ATLASGAL sources measured in the fourth quadrant \citep{2018A&A...609A.125W} using the Parkes telescope. This resulted in an ATLASGAL subsample of 264 clumps detected in NH$_2$D and NH$_3$ within a FWHM beamwidth of the Parkes telescope of 60$\arcsec$, slightly larger than the FWHM beamwidth of the IRAM telescope of 29$\arcsec$ at $\sim$ 86 GHz. The NH$_2$D line widths are compared with the NH$_3$ (1,1) line widths obtained from hyperfine structure fits in Fig. \ref{dvnh2d-dv11}, where the sources with an error in the optical depth smaller than 50\% are shown as red triangles and the clumps with an error in the optical depth larger than 50\% as black points. The straight line indicates equal line widths. The whole sample is distributed equally around the straight line and this hints at a correlation between the NH$_2$D and NH$_3$ line width within the noise. Figure \ref{dvnh2d-dv11} suggests that the NH$_2$D line at 86 GHz and NH$_3$ (1,1) line therefore trace similar regions within a source. Although the critical density of the NH$_2$D transition at 86 GHz is about a factor 50 higher than that of the NH$_3$ lines (see Table \ref{spectroscopic}), studies of high-mass star-forming regions also reveal an approximate spatial correlation between the emission from the two molecules \citep{2010AA...517L...6B,2011A&A...530A.118P}. The median NH$_2$D line width of $\sim 2$ km~s$^{-1}$ agrees with the average NH$_3$ (1,1) line width measured for the whole ATLASGAL sample in \cite{2012A&A...544A.146W}. The red contour lines of the sources with an error in the optical depth of less than 50\% indicate slightly smaller NH$_2$D than NH$_3$ line width.

\subsection{NH$_2$D column density}
\label{deuteration calculation}
The total column density of ortho and para NH$_2$D is derived assuming that the energy levels are in LTE, that is, that they are populated according to a Boltzmann distribution. Non-LTE conditions are pointed out in Sect. \ref{nh2d excitation}.
To calculate the column density we distinguish between subsamples with and without detected hyperfine structure. For clumps without detected hyperfine structure we mark the optical depth given in Table \ref{parline-nh2d85} with a star, while we give the optical depth of sources with detected hyperfine structure without a star. For the frequency of the NH$_2$D line at $\sim 86$ GHz we use the Rayleigh-Jeans approximation for a mean kinetic temperature of our sample at 20 K. We calculate the source-averaged column density of sources, for which the hyperfine structure components are detected and their ratio provides a measurement of the optical depth. For these clumps, the optical depth is well determined and the column density is derived from the lower energy level by
\begin{eqnarray}
 N_{\rm l} = \frac{8 \pi k}{c^3 h} \frac{g_{\rm l}}{g_{\rm u}} A^{-1} \nu^2 T_{\rm ex} \tau \Delta \rm v,
 \end{eqnarray}
 with the statistical weight of the upper and lower levels, $g_{\rm u}$ and $g_{\rm l}$, respectively, the Einstein $A$ coefficient in s$^{-1}$, the frequency of the NH$_2$D transition at 86 GHz, $\nu$, in GHz, the excitation temperature, $T_{\rm ex}$, in K, the optical depth of the NH$_2$D transition, $\tau$, and the NH$_2$D FWHM line width, $\Delta$v, in km~s$^{-1}$. We use the NH$_3$ kinetic temperature given in \cite{2012A&A...544A.146W} and \cite{2018A&A...609A.125W} as excitation temperature assuming that NH$_2$D and NH$_3$ are co-spatial, as indicated in Fig. \ref{dvnh2d-dv11}, and therefore have similar gas temperatures. For sources with detected hyperfine structure we calculate the total NH$_2$D column density in cm$^{-2}$ from the optical depth and kinetic temperature, which depend on line ratios, and we therefore compute a source-averaged quantity by  
\begin{eqnarray}\label{nh2d column density}
 N_{\rm tot} = 1.94 \times 10^3 \nu^2 A^{-1} \frac{Q (T_{\rm ex})}{g_{\rm u}} {\rm exp} \left( \frac{E_{\rm u}}{T_{\rm ex}} \right) T_{\rm ex} \tau \Delta \rm v, 
\end{eqnarray}
where $Q$ is the partition function and $E_{\rm u}$ the upper energy level in K (see Table \ref{spectroscopic}). Clumps without detected hyperfine structure and insufficient S/N to reliably determine an optical depth have optical depths with an error of greater than 50\% in Table \ref{parline-nh2d85}. We use the integrated intensity of the NH$_2$D line at 86 GHz, $\int T_{\rm mb} \rm dv$, in K km~s$^{-1}$, that is a measure derived over the whole beam, to calculate the beam-averaged column density in the optically thin case,
\begin{eqnarray}\label{col-opt-thin}
N_{\rm tot} = 1.94 \times 10^3 \nu^2 A^{-1} \frac{T_{\rm ex} \int T_{\rm mb} \ \rm dv}{T_{\rm ex} - 2.7 \ \rm K} {\rm exp} \left( \frac{ E_{\rm u}}{T_{\rm ex}} \right) \frac{Q (T_{\rm ex})}{g};
\end{eqnarray}
we derive Q from a fit to the partition functions measured for different rotation temperatures in the range between 9 and 300 K that was taken from the Cologne Database for Molecular Spectroscopy (CDMS)\footnote{see https://www.astro.uni-koeln.de/cdms}. The measured values of the partition functions listed on the CDMS are determined from the sum of the population of the 86 and 110 GHz transitions and take the spin multiplicity of the $^{14}$N nucleus, $g_I = 3$, into account. The fit yields a total partition function for ortho and para NH$_2$D of $Q = 1.04 \ T_{\rm ex}^{1.41}$. We obtain the statistical weight from $g = g_{J} \times g_{I} = 27$ with the angular momentum J = 1, the N and D\footnote{Because we work with the partition function and molecular line parameters from CDMS, the D nuclear spin is not taken into account.} nuclear spins of 1, and the H nuclear spin of 1/2.

\begin{table*}
\begin{minipage}{\textwidth}
\caption[NH$_3$ and NH$_2$D column densities and deuteration]{Source-averaged NH$_3$ and ortho NH$_2$D column densities and deuteration. Errors are given in parentheses. The full table is available at the CDS.}              
\label{parline-nh2d-nh3}      
\centering                                      
\begin{tabular}{l c c c c c}          
\hline\hline                        
 & $N_{\rm NH_3}$ & $\eta$ & $N_{\rm NH_2D}$ (86 GHz) & [NH$_2$D]/[NH$_3$] & $\int T_{\rm mb} \ \rm dv$ (86 GHz) \\ 
Name  & (10$^{15}$ cm$^{-2}$) &  & (10$^{13}$ cm$^{-2}$) &  & (K km~s$^{-1}$) \\                    
\hline    
%Added by TeX Support                              
 G10.62-0.42 & 2.33  $(\pm$0.16)  & 0.13 $(\pm$0.02) & 59.95 $(\pm$23.91) & 0.26 $(\pm$0.10) & 1.5 $(\pm$0.1) \\
G27.37-0.17 & 3.21 $(\pm$0.15)  & 0.17 $(\pm$0.01) & 172.19 $(\pm$46.32) & 0.53 $(\pm$0.15) & 4.9 $(\pm$0.1) \\
G30.42-0.23 & 3.72 $(\pm$0.19) & 0.14 $(\pm$0.01) & 160.16 $(\pm$36.57) & 0.43 $(\pm$0.10) & 4.3 $(\pm$0.1) \\
G30.79+0.20 &3.87 $(\pm$0.16) & 0.12 $(\pm$0.01) & 2.38 $(\pm$0.33) & 0.05 $(\pm$0.08) & 1.6 $(\pm$0.1) \\
G31.58+0.08 & 1.99 $(\pm$0.21)  & 0.1 $(\pm$0.01) & 0.74 $(\pm$0.25) & 0.04 $(\pm$0.02) & 0.4 $(\pm$0.1) \\
G34.26+0.15 & 4.28 $(\pm$0.47) & 0.16 $(\pm$0.13) & 3.94 $(\pm$0.87) & 0.03 $(\pm$0.02) & 2.1 $(\pm$0.1) \\
G48.99-0.30 & 2.89 $(\pm$0.20) & 0.12 $(\pm$0.01) & 5.15 $(\pm$0.56) & 0.15 $(\pm$0.02) & 2.9 $(\pm$0.1) \\
G49.27-0.34 & 3.43 $(\pm$0.20) & 0.15 $(\pm$0.01) & 7.1 $(\pm$0.97) & 0.14 $(\pm$0.02) & 4.4 $(\pm$0.2) \\
G49.40-0.21 & 1.86 $(\pm$0.14) & 0.17 $(\pm$0.02) & 64.67 $(\pm$21.92) & 0.35 $(\pm$0.12) & 2.3 $(\pm$0.1) \\
G58.47+0.43 & 1.45 $(\pm$0.15) & 0.15 $(\pm$0.02) & 1.73 $(\pm$0.33) & 0.08 $(\pm$0.02) & 1.1 $(\pm$0.1) \\
G305.23-0.02 & 2.04 $(\pm$0.64) & 0.02 $(\pm$0.01) & 1.14 $(\pm$0.56) & 0.25 $(\pm$0.18) & 0.8 $(\pm$0.3) \\
G305.82-0.11 & 1.69 $(\pm$0.35) & 0.05 $(\pm$0.01) & 1.69 $(\pm$0.71) & 0.22 $(\pm$0.12) & 1.2 $(\pm$0.2) \\
G309.38-0.13 & 1.79 $(\pm$0.39) & 0.05 $(\pm$0.01) & 1.46 $(\pm$0.45) & 0.17 $(\pm$0.08) & 1.0 $(\pm$0.2) \\
G310.01+0.39 & 1.86 $(\pm$0.43) & 0.03 $(\pm$0.01) & 1.68 $(\pm$0.57) & 0.36 $(\pm$0.19) & 1.1 $(\pm$0.3) \\
G351.74-0.58 & 3.85 $(\pm$0.17) & 0.12 $(\pm$0.01) & 98.78 $(\pm$31.59) & 0.25 $(\pm$0.08) & 3.5 $(\pm$0.2) \\
G351.78-0.52 & 3.66 $(\pm$0.11) & 0.19 $(\pm$0.01) & 125.34 $(\pm$26.70) & 0.35 $(\pm$0.07) & 4.4 $(\pm$0.2) \\
G353.41-0.36 & 1.91 $(\pm$0.09) &  0.19 $(\pm$0.02) & 128.84 $(\pm$39.46) & 0.68 $(\pm$0.21) & 8.6 $(\pm$0.2) \\

\hline                                            
\end{tabular}
\end{minipage}
\end{table*}

\subsection{NH$_2$D excitation}
\label{nh2d excitation}
Furthermore, we test the assumption that the NH$_3$ rotational temperature between the (1,1) and (2,2) inversion transition is equal to the NH$_2$D excitation temperature using the NH$_2$D 1$_{11}-1_{01}$ line at 110 GHz and the NH$_2$D $2_{12}-2_{02}$ transition at 74 GHz that was observed towards a subsample of 24 ATLASGAL sources. The rotational temperature between the NH$_2$D transitions at 110 GHz and 74 GHz is determined from the simultaneous modelling of the 74 GHz and 110 GHz lines (see Fig. \ref{nh2d-lines-model}) using MCWeeds \citep{2017A&A...603A..33G}. This assumes equal excitation temperatures of the 1$_{11}-1_{01}$  and $2_{12}-2_{02}$ lines under LTE conditions. This package is based on WEEDS \citep{2011A&A...526A..47M} from the CLASS software and adds Bayesian statistics and fitting algorithms to Weeds to automatise the simultaneous fitting of the lines of several species. Furthermore, errors on the NH$_2$D temperature are also estimated by MCWeeds. It determines the NH$_2$D rotational temperature for given starting values of the input parameters, that is, the NH$_2$D column density, temperature, line width, and source size. As starting values we used the results from our hyperfine structure fitting of the NH$_2$D transition at 110 GHz and the NH$_3$(1,1) and (2,2) lines. The fit parameters, the radial velocity of para NH$_2$D at 74 GHz, $\rm v_{LSR_{74}}$, the line width, $\Delta \rm v_{74}$, and the NH$_2$D temperature, $T_{\rm rot_{74}}$, are given in Table \ref{parline-nh2d74}.

\begin{figure}[h]
\centering
\includegraphics[angle=0,width=9.0cm]{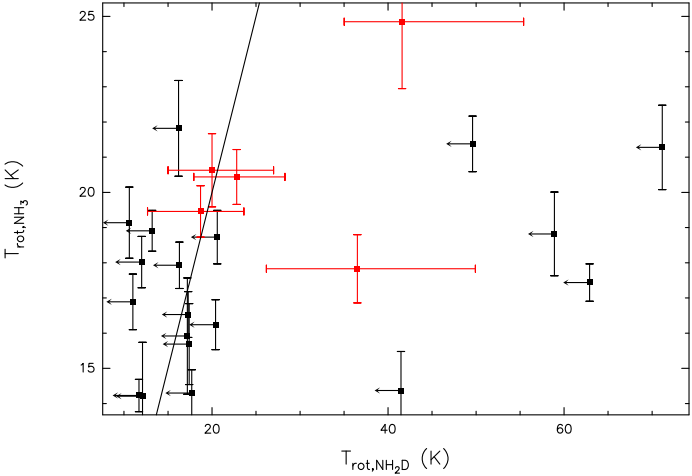}
\caption[Comparison of NH$_2$D and NH$_3$ rotational temperature]{Observed NH$_3$ rotational temperature between the (1,1) and (2,2) inversion transition is shown against modelled NH$_2$D rotational temperature between the 1$_{11}-1_{01}$ line at 110 GHz and the $2_{12}-2_{02}$ transition at 74 GHz. Sources with a detected 74 GHz line are marked in red, and non-detections with an upper limit of the NH$_2$D temperature are labelled as black arrows. The straight line corresponds to equal temperatures.}\label{trot-nh2d-nh3}
\end{figure}

While modelling of the two NH$_2$D lines leads to the NH$_2$D temperature for sources with a detected NH$_2$D transition at 74 GHz, we obtained an upper limit to the NH$_2$D temperature for the non-detections. Comparison of the rotational temperature between the NH$_3$ (1,1) and (2,2) inversion transition \citep{2012A&A...544A.146W} with the NH$_2$D temperature in Fig. \ref{trot-nh2d-nh3} yields a difference in the NH$_2$D and NH$_3$ rotational temperatures for a subsample of the sources. However, a fraction of the clumps exhibit larger NH$_3$ than NH$_2$D rotational temperatures, in some cases supported by upper limits to the NH$_2$D temperature. We note that the critical density is proportional to the Einstein A-coefficient, which in turn is proportional to $\nu^3$. A factor 4.6 higher frequency of NH$_2$D at 110 GHz than of the NH$_3$(1,1) and (2,2) lines leads to a much higher critical density of NH$_2$D than NH$_3$ (see Table \ref{spectroscopic}). 

In addition to our assumption of the same excitation temperature for the 74 GHz and 110 GHz lines as there would be for LTE, we also run a non-LTE modelling using RADEX \citep{2007A&A...468..627V} with a fixed kinetic temperature of 20 K and a fixed NH$_2$D column density of $1.6 \times 10^{15}$ cm$^{-2}$ corresponding to the mean values of our ATLASGAL sample. We use the recent calculations by \cite{2014MNRAS.444.2544D} for the collisional rates with the H$_2$ molecule for our non-LTE calculations. This resulted in similar excitation temperatures of the 74 GHz and 110 GHz NH$_2$D lines at high densities ($> 10^6$ cm$^{-3}$). The five detections at 74 GHz from our sample are classified as young stellar objects (YSOs), hot cores, or HII regions. Urquhart et al. (in prep.) determined volume densities of $\sim 10^6$ cm$^{-3}$ based on the dust continuum from the ATLASGAL survey for these sources, which are also consistent with the typical density of YSOs probed in NH$_2$D by \cite{2011A&A...530A.118P}.
 
We compared NH$_2$D column densities using the NH$_3$ rotational temperature and NH$_2$D temperature as excitation temperature for sources with detections at 74 and 110 GHz. This shows that for the subsample with larger NH$_3$ rotational temperatures than NH$_2$D temperatures we overestimate the NH$_3$ deuteration by 47\%. Similarly, we underestimate the NH$_3$ deuteration by $\sim$32\% for the subsample with lower NH$_3$ rotational temperature than the NH$_2$D temperature. We cannot exclude subthermal excitation due to densities below the NH$_2$D line critical densities that lead to lower temperatures.

\begin{figure}[h]
\centering
\includegraphics[angle=0,width=9.0cm]{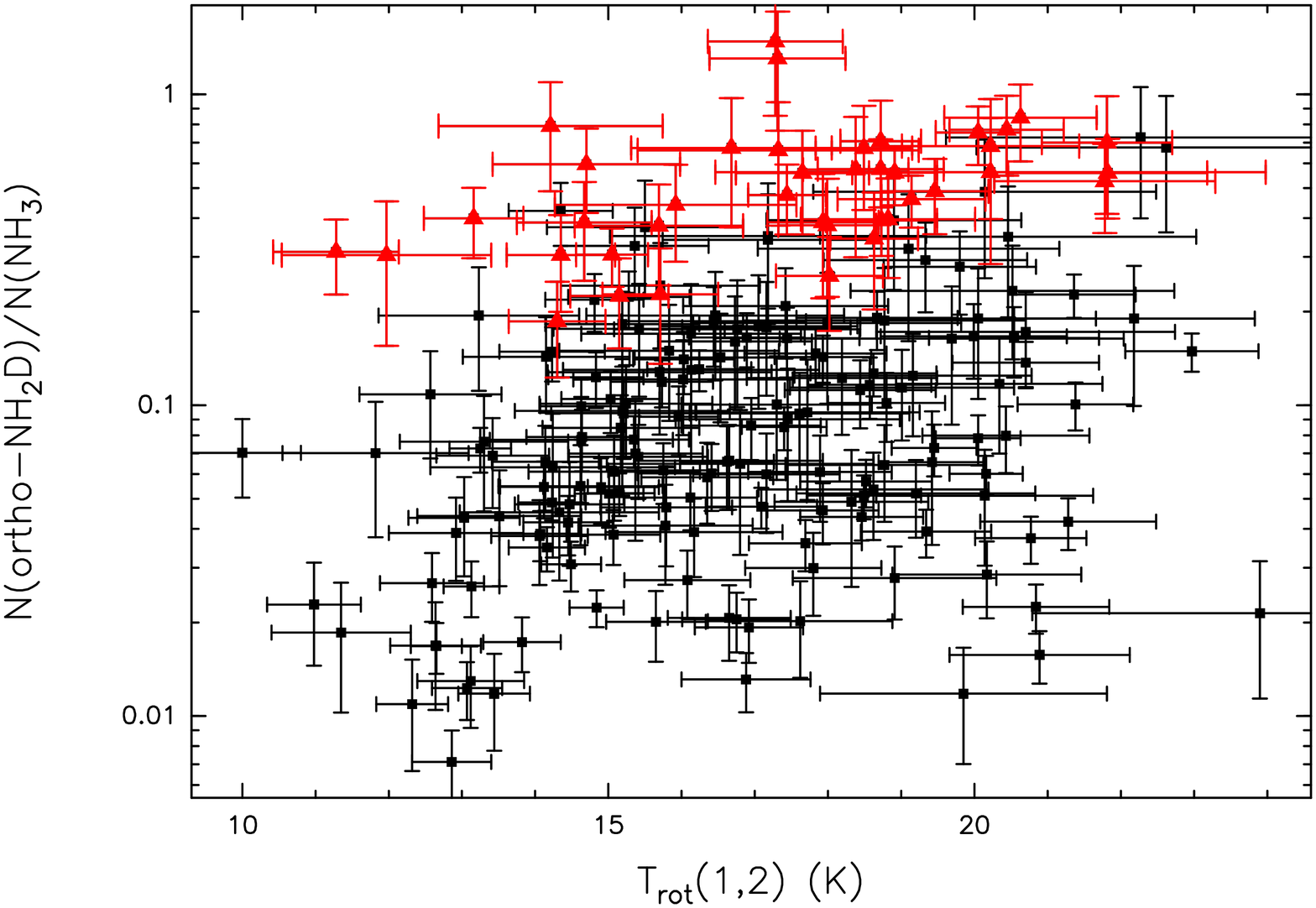}\vspace*{0.5cm}
\includegraphics[angle=0,width=9.0cm]{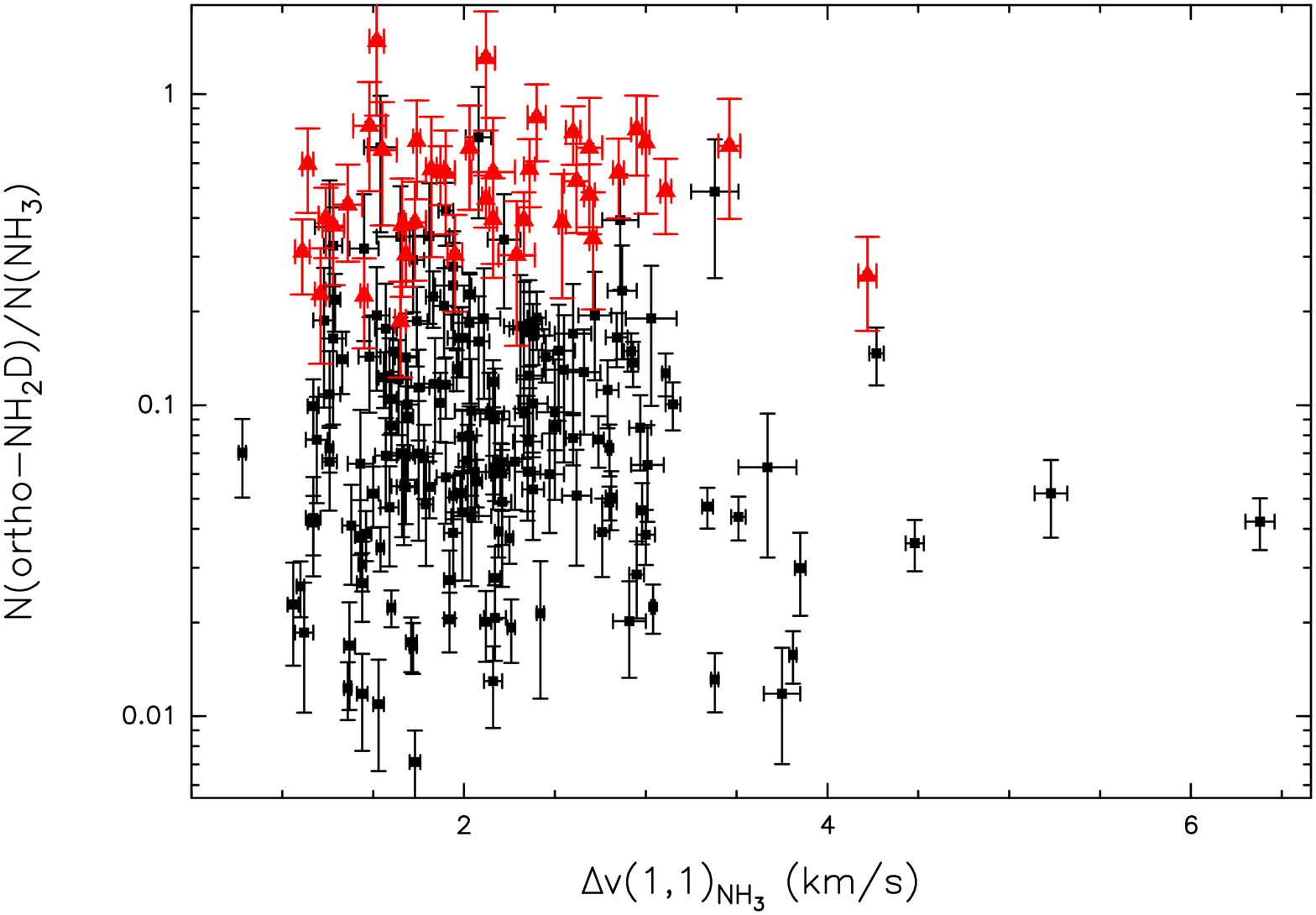}\vspace*{0.5cm}
\includegraphics[angle=0,width=9.0cm]{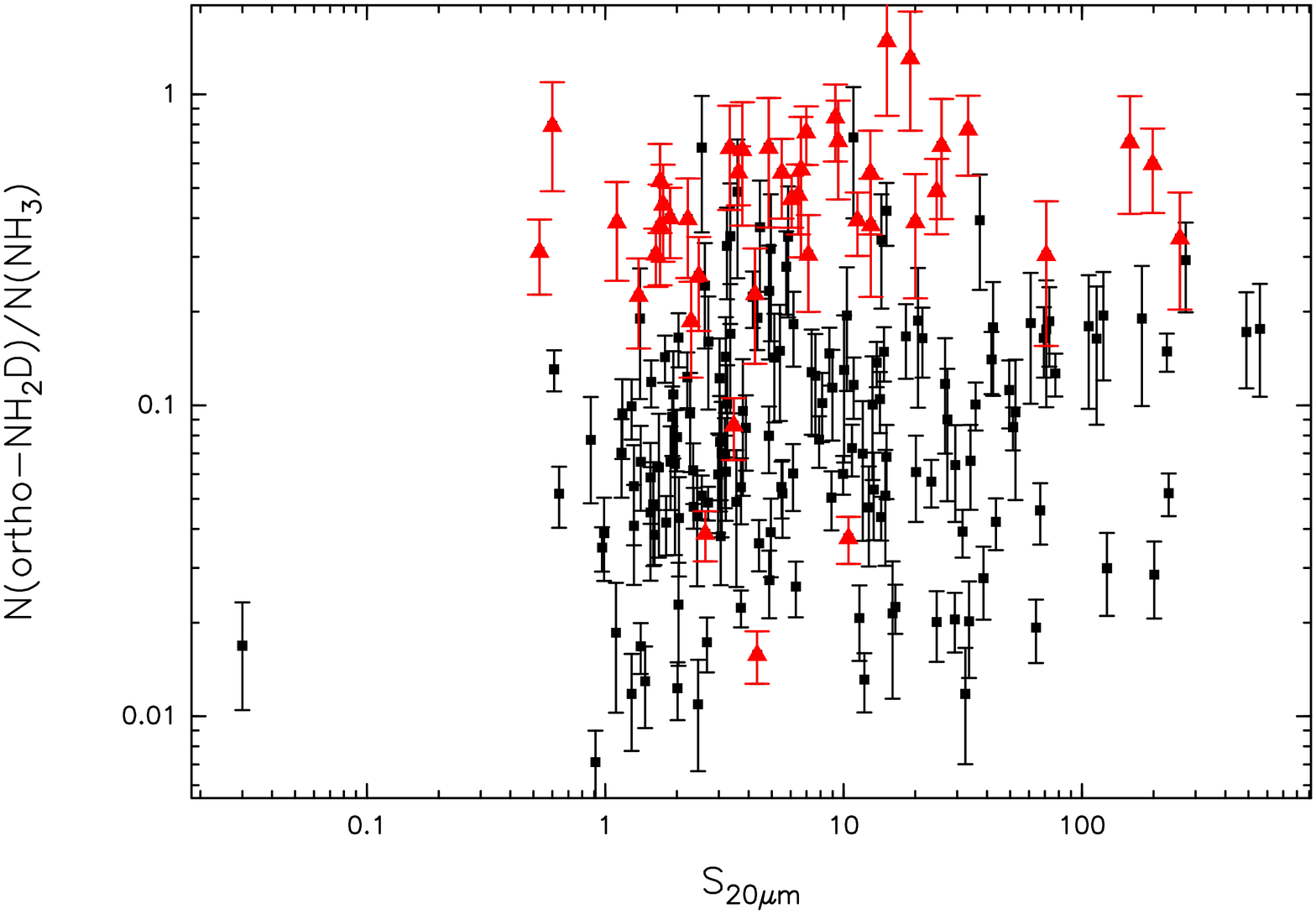}
\caption[Comparison of ortho NH$_2$D/NH$_3$ column density ratio with NH$_3$ rotational temperature, NH$_3$ linewidth and MSX 21 $\mu$m flux]{NH$_2$D to NH$_3$ column density ratio compared with the NH$_3$ rotational temperature, line width, and the MSX 21 $\mu$m flux for the ATLASGAL sources; clumps with and without detected hyperfine structure are shown with red triangles and  black points, respectively.}\label{nh2d-dv11-trot}
\end{figure}

\subsection{NH$_3$ deuteration}
\label{nh3 deuteration}
We determine the fractionation ratio by the total column density of ortho and para NH$_2$D (see Sect. \ref{deuteration calculation}) to the NH$_3$ column density ratio assuming that the two molecules originate from the same gas. The source-averaged NH$_3$ column density was calculated for ATLASGAL sources in the first quadrant in \cite{2012A&A...544A.146W} and for clumps in the fourth quadrant in \cite{2018A&A...609A.125W}. To account for the different column density determinations of NH$_2$D (see Sect. \ref{deuteration calculation} source- vs. beam-averaged) we derive the NH$_3$ deuteration in two ways. We divide the source-averaged NH$_2$D and NH$_3$ column densities for the NH$_2$D sources with detected hyperfine structure. In the optically thin case we correct the NH$_2$D column density for the beam dilution using the beam filling factor derived from NH$_3$ observations \citep[see Sect. 4.5 in][]{2012A&A...544A.146W} to estimate the source-averaged NH$_2$D to NH$_3$ column density ratio. Table \ref{parline-nh2d-nh3} shows the source-averaged NH$_3$ column density, the beam filling factor, the NH$_2$D column density, the NH$_2$D to NH$_3$ column density ratio, and the integrated intensity of the NH$_2$D line at 86 GHz, $\int T_{\rm mb} \ \rm dv$, with their errors. We measure [NH$_2$D]/[NH$_3$] ratios between 0.007 and 1.6. The distribution of the deuterium fraction of NH$_3$ is shown as a function of the rotational temperature in the upper panel of Fig. \ref{nh2d-dv11-trot}. Optically thin fits result in lower limits, because the NH$_2$D optical depth is not known for sources without detected hyperfine structure and we cannot exclude a considerable optical depth. We might therefore underestimate the column density by using equation \ref{col-opt-thin} in the optically thin case. However, there is still a range of one order of magnitude in the fractionation ratio. We consider only measurements with relative errors on the presented NH$_2$D line parameters of less than $\sim$ 50\% for the correlation plots. In 109 out of the 264 clumps detected in NH$_2$D and NH$_3$ (41\%) with reliable hyperfine structure fits we determine high deuteration ($>$ 0.16) with 0.16 being the lowest deuteration in this sample. Of those, 18 sources exhibit errors in the [NH$_2$D]/[NH$_3$] ratio of greater than 50\%.

\subsection{The ortho-to-para ratio of NH$_2$D}
\label{ortho-to-para ratio}
\begin{figure}[h]
\centering
\includegraphics[angle=0,width=9.0cm]{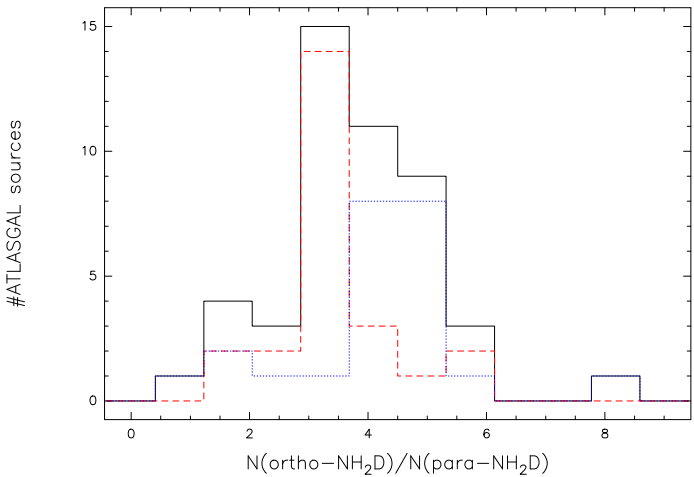}
\caption[Histogram of ortho-to-para NH$_2$D  column density ratio]{Number distribution of the column density ratio derived from the ortho and para NH$_2$D transitions shown in black for the subsample detected in NH$_2$D with a NH$_3$ counterpart, as dashed red line for ATLASGAL sources with a NH$_3$ deuteration $< 0.2$ and as dotted blue line for a deuterated factor $> 0.2$.}\label{ortho-para-histo}
\end{figure}

For a subsample of 113 ATLASGAL sources we detect the NH$_2$D ortho line at 86 and para line at 110 GHz. We calculate the ortho-to-para column density ratio assuming the same beam filling for the two transitions, as expected for unresolved, clumpy structure observed within molecular clouds that fill the beam \citep{1990ApJ...356..513S,1985A&A...152..371P}. 
We distinguish between three different cases depending on the optical depth of the source: We divide the source-averaged column densities derived from the transition at 86 and 110 GHz as given in equation \ref{nh2d column density} with the molecular line parameters listed in Table \ref{spectroscopic} for sources with detected hyperfine structure in ortho and para NH$_2$D. Here, Q is determined from separate fits to the partition functions of ortho and para NH$_2$D at 86 and 110 GHz against rotation temperatures from 3 to 300 K published on the CDMS. These result in  $Q = 0.59 \ T_{\rm ex}^{1.49}$ for the ortho line and  $Q = 0.19 \ T_{\rm ex}^{1.49}$ for the para line. The ortho and para partition functions differ by a factor three which results from the spin multiplicity. Our fit of the two partition functions yields an exponent of 3/2, as expected for slightly asymmetric top molecules \citep{2015PASP..127..266M}.\\
For sources without detected hyperfine structure in ortho and para NH$_2$D we derive the beam-averaged NH$_2$D column densities using equation \ref{col-opt-thin} with the observed intensities and molecular line parameters for the transitions at 86 and 110 GHz given in Table \ref{spectroscopic} and separate ortho and para partition functions. To correct the NH$_2$D column densities for the beam dilution we divide the two by the beam filling factor; they cancel out in the computation of the ortho-to-para ratio from the column density ratio.\\
As we cannot detect the hyperfine structure of the NH$_2$D line at 110 GHz for the majority of sources, we cannot measure the optical depth at 110 GHz directly. To derive the ortho-to-para ratio for the subsample with detected hyperfine structure in ortho NH$_2$D we calculated the column density of para NH$_2$D in the optically thin approximation from equation \ref{col-opt-thin} using the molecular line parameters at 110 GHz and the para partition function. We get an estimate of the optical depth at 110 GHz iteratively from the known optical depth at 86 GHz: We start with an ortho-to-para ratio of three as expected from their statistical values, calculate the optical depth at 110 GHz by the ratio of the optical depth at 86 GHz to the ortho-to-para ratio, and compute the column density ratio determined from the lines at 86 and 110 GHz. This is used subsequently as the ortho-to-para ratio in the calculation of the optical depth at 110 GHz until the ortho-to-para ratio converges. With the resulting optical depth at 110 GHz we multiply the column density of para NH$_2$D by the factor $\tau'/(1-{\rm exp}(-\tau'))$ with $\tau' = 0.679 \ \tau^{0.911}$ , which corrects for the optical depth of the line at 110 GHz \citep{1989ApJ...340L..37S}, and divide by the beam filling factor. The ortho-to-para ratio is then determined from the source-averaged ortho and para column densities. We also tested if the ortho-to-para ratio depends on the initial value chosen for this ratio. We therefore varied this from 3 to 1 and 5, but the iteratively determined ortho-to-para ratios of the whole sample did not change statistically. \\
Using an initial ortho-to-para ratio of 3 we obtain the distribution of the iteratively derived ortho-to-para ratios shown by the black histogram in Fig. \ref{ortho-para-histo} with a median ortho-to-para ratio of 3.7 and a standard deviation of 1.2. Our ortho-to-para ratio is close to the value expected from the nuclear statistical weights of 3 and 1 for the ortho and para NH$_2$D species, respectively. The column density derived from the NH$_2$D transition at 110 GHz and the ortho-to-para ratio with their errors are given in Table \ref{parline-nh2d-110}.

\begin{table*}
\begin{minipage}{\textwidth}
\caption[Column density derived from the NH$_2$D para line and ortho-to-para ratio]{Column density derived from the NH$_2$D line at 110 GHz and ortho-to-para ratio. Errors are given in parentheses. The full table is available at the CDS.}             
\label{parline-nh2d-110}     
\centering                                     
\begin{tabular}{l c c}          
\hline\hline                        
 &  N$_{\rm NH_2D}$ (110 GHz) & $N_{\rm ortho} ({\rm 86 GHz})/N_{\rm para} ({\rm 110 GHz})$ \\ 
Name  &  (10$^{13}$ cm$^{-2}$)  &   \\                    
\hline    
%Added by TeX Support                              
 G10.21-0.30 & 0.19 $(\pm$0.10) & 6.0 $(\pm$3.3) \\
G10.67-0.22 & 0.37 $(\pm$0.10) & 3.0 $(\pm$0.9) \\
G11.92-0.61 & 0.82 $(\pm$0.14) & 4.3 $(\pm$1.9) \\
G11.94-0.61 & 0.59 $(\pm$0.13) & 3.9 $(\pm$1.8) \\
G12.50-0.22 & 0.78 $(\pm$0.13) & 3.2 $(\pm$1.2) \\
G13.90-0.51 & 0.60 $(\pm$0.12) & 4.4 $(\pm$1.3) \\
G14.18-0.53 & 1.70 $(\pm$0.31) & 1.3 $(\pm$0.4) \\
G14.23-0.51 & 1.73 $(\pm$0.19) & 2.9 $(\pm$0.4) \\
G14.25+0.07 & 0.48 $(\pm$0.17) & 3.9 $(\pm$1.6) \\
G16.58-0.05 & 0.43 $(\pm$0.13) & 3.5 $(\pm$1.3) \\
G17.65+0.17 & 0.68 $(\pm$0.17) & 3.4 $(\pm$1.4) \\
G22.04+0.22 & 0.55 $(\pm$0.17) & 3.2 $(\pm$1.2) \\
G30.42-0.23 & 1.24 $(\pm$0.16) & 4.9 $(\pm$1.3) \\
G48.99-0.30 & 1.28 $(\pm$0.17) & 3.0 $(\pm$0.5) \\
G49.27-0.34 & 1.32 $(\pm$0.21) & 4.0 $(\pm$0.8) \\
G58.47+0.43 & 0.36 $(\pm$0.09) & 3.5 $(\pm$1.1) \\

\hline                                             
\end{tabular}
\end{minipage}
\end{table*}

\section{Discussion}
\label{discussion}
\subsection{NH$_3$ and NH$_2$D rotational temperature}
The NH$_3$ rotational temperature between the (1,1) and (2,2) inversion transition is compared with the rotational temperature between the NH$_2$D transitions at 110 GHz and 74 GHz in Fig. \ref{trot-nh2d-nh3}. This comparison reveals a large range of variation in temperature; two clumps, G31.41+0.31 and G34.26+0.15, exhibit extremely high NH$_2$D temperatures of 37 and 42 K. The high excitation temperature points to the hot molecular core as the main origin of the deuterated ammonia in these sources. We found little deuterium fractionation for G31.41+0.31 of 0.15 and for G34.26+0.15 of 0.02 as expected at the high temperatures of hot cores and close to the values reached by models of hot molecular cores.

High-resolution observations of high-mass star-forming regions in NH$_2$D have shown that while NH$_2$D shows an excellent correlation with dust continuum in high-mass cold cores, deuterated ammonia avoids the dense peaks close to very luminous protostars \citep{2010A&A...517L...6B,2011A&A...530A.118P}. While this might be a temperature effect in regions with complex dynamics, without further high-angular-resolution observations, we are unable to confirm that the major contributing factor is deuteration.

\subsection{Comparison of NH$_3$ deuteration in ATLASGAL sources with other samples}
\label{other samples}
We derive a deuterium fraction between 0.007 and 1.6 in Sect. \ref{nh3 deuteration}, sources without detected hyperfine structure exhibit low [NH$_2$D]/[NH$_3$] ratios with a median of $0.05 \pm 0.01,$ and clumps with detected hyperfine structure in NH$_2$D and NH$_3$ have a higher deuteration with a median of $0.4 \pm 0.05$. The [NH$_2$D]/[NH$_3$] ratios of approximately 1 are among the highest given in the literature so far. Other high average deuteration factors were estimated to be 0.8 for starless cores associated with the UCHIIR IRAS 20293+3952 \citep{2010AA...517L...6B}, 0.66 was derived for a small sample of clumps embedded in infrared-dark clouds by \cite{2007AA...467..207P}, and 0.5 in the low-mass starless core L1544 in Taurus \citep{2007AA...470..221C}. 

A lower deuteration of 0.33 was obtained by \cite{2003AA...403L..25H} towards low-mass protostellar cores in Perseus, where optical depth could be measured and [NH$_2$D]/[NH$_3$] ratios best determined, and a deuteration of 0.1 was found in dark clouds by \cite{2000AA...356.1039T}. \cite{2001ApJ...554..933S} observed NH$_2$D in low-mass and quiescent protostellar cores and measured a deuteration factor between 0.003 and 0.13, and \cite{2000ApJ...535..227S} determined a similar range in deuteration in dark cloud cores located mostly in the Taurus and Ophiuchus regions. A comparison of the deuterium fraction of NH$_3$ towards the low-mass samples shows that \cite{2000ApJ...535..227S} and \cite{2001ApJ...554..933S} estimated lower values on average than \cite{2003AA...403L..25H}. This latter author suggested an increase in deuteration from larger to smaller scales towards protostellar cores as reason for the low [NH$_2$D]/[NH$_3$] ratio obtained by \cite{2001ApJ...554..933S} and \cite{2000ApJ...535..227S}, who measured NH$_2$D with a larger beam width than \cite{2003AA...403L..25H} or with an offset from the dust peak. We summarise the deuterium fractionation of the different samples from the literature in Table \ref{deuteration-literature} with the source number, the NH$_2$D detection rate, the NH$_2$D to NH$_3$ column density ratio, and reference. Most of these studies calculated the NH$_2$D column density under the assumption of LTE; only \cite{2007AA...470..221C} and \cite{2001ApJ...554..933S} used non-LTE models.

For comparison of the NH$_3$ deuteration in the ATLASGAL sample other studies with similar beamwidth as our NH$_3$ and NH$_2$D measurements are available in the literature: \cite{2015A&A...575A..87F} observed the two molecules in dense cores associated with different evolutionary phases of high-mass star formation and determined a deuterium fraction of NH$_3$ between 0.21 and 0.34. \cite{2007AA...467..207P} found a deuterium fractionation from 0.004 to 0.7 that is similar for most clumps embedded in infrared-dark clouds. These [NH$_2$D]/[NH$_3$] ratios are consistent with those of the ATLASGAL sources, although we estimate an even higher deuteration than \cite{2007AA...467..207P}, up to 1.6 in a few ATLASGAL clumps. Recently, \cite{2019A&A...631A..63S} compared two approaches to model deuterium fractionation that differ in their mechanism to describe ion-molecule proton-donation reactions. The full scrambling model comprises multiple interchanges of atoms including for example proton hop and proton exchange \citep{2004JMoSp.228..635O}. The time evolution of their full scrambling model \citep{2015A&A...581A.122S} leads to a deuteration of the order of 10$^{-1}$ over one phase, similar to models of earlier studies \citep{2000A&A...361..388R,2002P&SS...50.1189M}. To reach such a high rate as measured towards the ATLASGAL sample we speculate that each of several clumps within the beam goes through a cycle of enrichment consisting of freeze-out and evaporation during a dense and cold phase. The high deuterium fractionation of our sample might then result from the accumulation of [NH$_2$D]/[NH$_3$] ratios of individual clumps within the beam.    NH$_3$ deuteration as high as our observed values is obtained by the model of \cite{2019A&A...631A..63S} which limits proton-donation reactions to proceed only through proton hop \citep[cf.][]{2018MNRAS.477.4454H}. This model predicts a [NH$_2$D]/[NH$_3$] ratio exceeding 1 for a density of 10$^6$ cm$^{-3}$ after one phase that lasts $\sim 10^5$ years.

A comparison of the [NH$_2$D]/[NH$_3$] ratio of the ATLASGAL sources with that of low-mass star-forming samples, for example 0.02 - 0.1 by \cite{2000AA...356.1039T}, 0.025 - 0.18 by \cite{2000ApJ...535..227S}, and 10$^{-3}$ - 10$^{-1}$ by \cite{2001ApJ...554..933S}, indicates at least similar deuteration in high-mass star-forming regions. \cite{2010ApJ...716..433K} found that at a certain radius cluster-forming clouds have more mass and therefore a higher density than their counterparts without cluster formation. As the timescale of deuteration has been found to be shorter with increasing density \citep{2017MNRAS.469.2602K}, we expect an enhanced deuterium fractionation of ATLASGAL sources resulting from the higher density of this high-mass star-forming sample.

\begin{table*}
\begin{minipage}{\textwidth}
\caption[NH$_2$D to NH$_3$ column densities]{Comparison of the NH$_3$ deuteration from ATLASGAL with other samples.}             
\label{deuteration-literature}    
\centering                                      
\begin{tabular}{c c c c c}          
\hline\hline                        
Sample selection & Sample size & detection rate in NH$_2$D (\%) & [NH$_2$D]/[NH$_3$]  & Reference  \\ 
\hline   
High-mass ATLASGAL clumps & 992 & 39 & 0.007 - 1.6 & this article \\
UCHIIR starless cores  &  7 & 100 &  $< 0.06$ - 0.8   &    \cite{2010AA...517L...6B}    \\  
Pre/protocluster clumps &  32 & 69   &     0.004 - 0.67 & \cite{2007AA...467..207P} \\
Pre-stellar core L 1544 & 1 & 100 & 0.5 & \cite{2007AA...470..221C}\tablefootmark{*} \\
Perseus protostellar cores & 7 & 100 & 0.17 - 0.33 & \cite{2003AA...403L..25H} \\
Dense cores & 2 & 100 & 0.02, 0.1 & \cite{2000AA...356.1039T} \\
Low-mass protostellar cores & 32 & 70 & 0.003 - 0.13 & \cite{2001ApJ...554..933S}\tablefootmark{*} \\
Dark molecular cloud cores & 16 & 50 & 0.025 -0.18 & \cite{2000ApJ...535..227S} \\
\hline                                            
\end{tabular}
\tablefoot{
        \tablefoottext{*}{Samples using non-LTE models for the determination of the NH$_2$D column density.}}
\end{minipage}
\end{table*}

\subsection{Does the deuterium fraction of NH$_3$ trace the evolution of ATLASGAL clumps?}
\textbf{Deuteration as an evolutionary tracer of low- and high-mass star formation:}
\cite{2010AA...517L...6B} detected NH$_2$D emission in a few starless cores, while it is not associated with YSOs in a high-mass star-forming region. The deuterium fraction of NH$_3$ therefore allowed them to distinguish between the pre-protostellar and protostellar phase. \cite{2011A&A...529L...7F} obtained differences in the deuteration of about $30$ cores at various evolutionary phases of high-mass star formation from the derivation of the [N$_2$D$^+$]/[N$_2$H$^+$] ratio. These latter authors found a decrease in the deuterium fraction from high-mass cores without stars to the evolutionary stages after the formation of protostars. In addition, they obtained a slight decrease of the fractionation with temperature and N$_2$H$^+$ line width. \cite{2015ApJ...814...31K} also found the narrowest line widths for the youngest Class 0 protostars in Orion with the largest amount of deuteration in H$_2$CO, which hints at early stages of star formation. However, they did not find a clear correlation of the [HDCO]/[H$_2$CO] ratio and the mass-to-luminosity ratio as a tracer of the evolutionary phase. \cite{2001ApJ...554..933S} claimed to have found a trend of increasing NH$_3$ deuteration with decreasing temperature. However, the sample of these latter authors contains only a few low-mass cores, which introduces a large statistical error. \cite{2009A&A...493...89E} determined the [N$_2$D$^+$]/N$_2$H$^+$] ratio for 20 protostellar cores in low-mass star-forming regions, which also yields an anticorrelation of deuteration with temperature.

While previous studies of deuteration focused on low-mass star-forming samples, molecules different from NH$_2$D, such as N$_2$D$^+$, and small source samples, in this section we examine whether or not the deuterium fraction of NH$_3$ is an indicator of the evolution in a large sample of high-mass star-forming regions. Because we obtained a statistically significant correlation between the NH$_3$ (1,1) line width and rotational temperature of ATLASGAL sources with the evolutionary phase in \cite{2012A&A...544A.146W}, we use these properties to investigate a dependence of the NH$_3$ deuteration on the evolution.

\textbf{NH$_3$ rotational temperature:} The [NH$_2$D]/[NH$_3$] ratio is plotted against the NH$_3$ rotational temperature between the (1,1) and (2,2) inversion transitions in the upper panel of Fig. \ref{nh2d-dv11-trot}. The rotational temperature of the ATLASGAL sources detected in NH$_2$D ranges between 10 and 24 K and NH$_3$ column densities between $1.6 \times 10^{14}$ and 10$^{16}$ cm$^{-2}$ (see \cite{2012A&A...544A.146W}). One expects the largest deuterium fraction of NH$_3$ at temperatures lower than 20 K resulting from collisions between H$_3^+$ and HD producing H$_2$D$^+$ and increasing the [H$_2$D$^+$]/[H$_3^+$] ratio \citep{2004A&A...427..887F,2000A&A...364..780R}. H$_2$D$^+$ is an important molecule in deuterium chemistry in cold clouds and enhances the deuteration of several other molecules. However, H$_2$D$^+$ is destroyed by neutral molecules such as CO at temperatures above $\sim 25$ K and by ortho-H$_2$ at temperatures below that as well. Freeze-out of CO onto dust grains increases the abundance of deuterated molecules at low temperatures and high densities. However, the deuteration of the ATLASGAL sample does not show an anticorrelation with temperature. This agrees with previous results from \cite{2011A&A...530A.118P} and \cite{2015A&A...575A..87F}. A small sample of IRDCs show no trend between [NH$_2$D]/[NH$_3$] ratio and temperature \citep{2011A&A...530A.118P} and the NH$_3$ deuteration does not rise with decreasing temperature for their sample of about 30 dense cores in different evolutionary phases of high-mass star formation \citep{2015A&A...575A..87F}. \\
Production of NH$_2$D begins during the early pre-protostellar phase. During the evolution of the ATLASGAL sample, an internal heating source forms in the innermost part of the clump and leads to a higher rotational temperature, while NH$_2$D is constantly accumulated in the still cold outer envelope. We speculate that this process might have the consequence of a constant deuteration for a rising rotational temperature. Alternatively, the luminosity of forming protostars within the clumps, for example G31.41+0.31 or G34.26+0.15, leads to an increase of temperature and evaporation and a subsequent increase of molecules whose deuteration has enhanced on grain surfaces resulting in an enhanced deuteration in the gas phase. The presence of clumps in a range of evolutionary phases within the beam that contain gas components with varying temperatures might also hide the dependence of increasing NH$_3$ deuteration on temperature.

\textbf{NH$_3$ line width:} We compared the fractionation ratio with the NH$_3$ line width as another evolutionary tracer in the middle panel of Fig. \ref{nh2d-dv11-trot}. The line widths range between 0.8 and 6.4 km~s$^{-1}$; there is no anticorrelation between deuteration and line width. This is in agreement with the deuteration of dense cores in massive star-forming regions \citep{2015A&A...575A..87F}, which does not depend on the line width either.

\textbf{MSX 21$\mu$m flux:} ATLASGAL sources without any star formation yet are not or only weakly detected at mid-infrared (MIR) or far-infrared (FIR) wavelengths, while clumps in a later evolutionary phase associated with a heating source emit at 21 $\mu$m. We therefore also compare the [NH$_2$D]/[NH$_3$] ratio with the MSX 21$\mu$m flux to investigate any trend of the NH$_3$ deuteration with evolution. However, the lower panel of Fig. \ref{nh2d-dv11-trot} shows a flat distribution of the MSX flux with the fractionation ratio.

In summary, Fig. \ref{nh2d-dv11-trot} illustrates that high-mass star-forming regions might be too complex to show a trend of decreasing NH$_3$ deuteration with increasing rotational temperature or line width. At the large distances of the ATLASGAL sample with a median distance of 4 kpc \citep{2015A&A...579A..91W} a clump likely harbours several cores at different evolutionary stages. The presence of multiple evolutionary phases within one source was also found by \cite{2014MNRAS.443.1555U}. While observations of nearby low-mass star-forming samples have a much higher spatial resolution and therefore focus on individual cores, the temperature and line width of an ATLASGAL clump results from an average of these properties over the beam width. An ATLASGAL source might therefore host cores with a large amount of NH$_3$ deuteration, low temperatures, and narrow line widths as well as warm, turbulent cores with broad line widths. However, averaging over the beam width results then in a high deuterium fraction of NH$_3$ at relatively high temperatures and broad line widths, and an overall constant distribution of the [NH$_2$D]/[NH$_3$] ratio over the range of NH$_3$ line parameter values.

\begin{figure}
\centering
\includegraphics[angle=0,width=9.0cm]{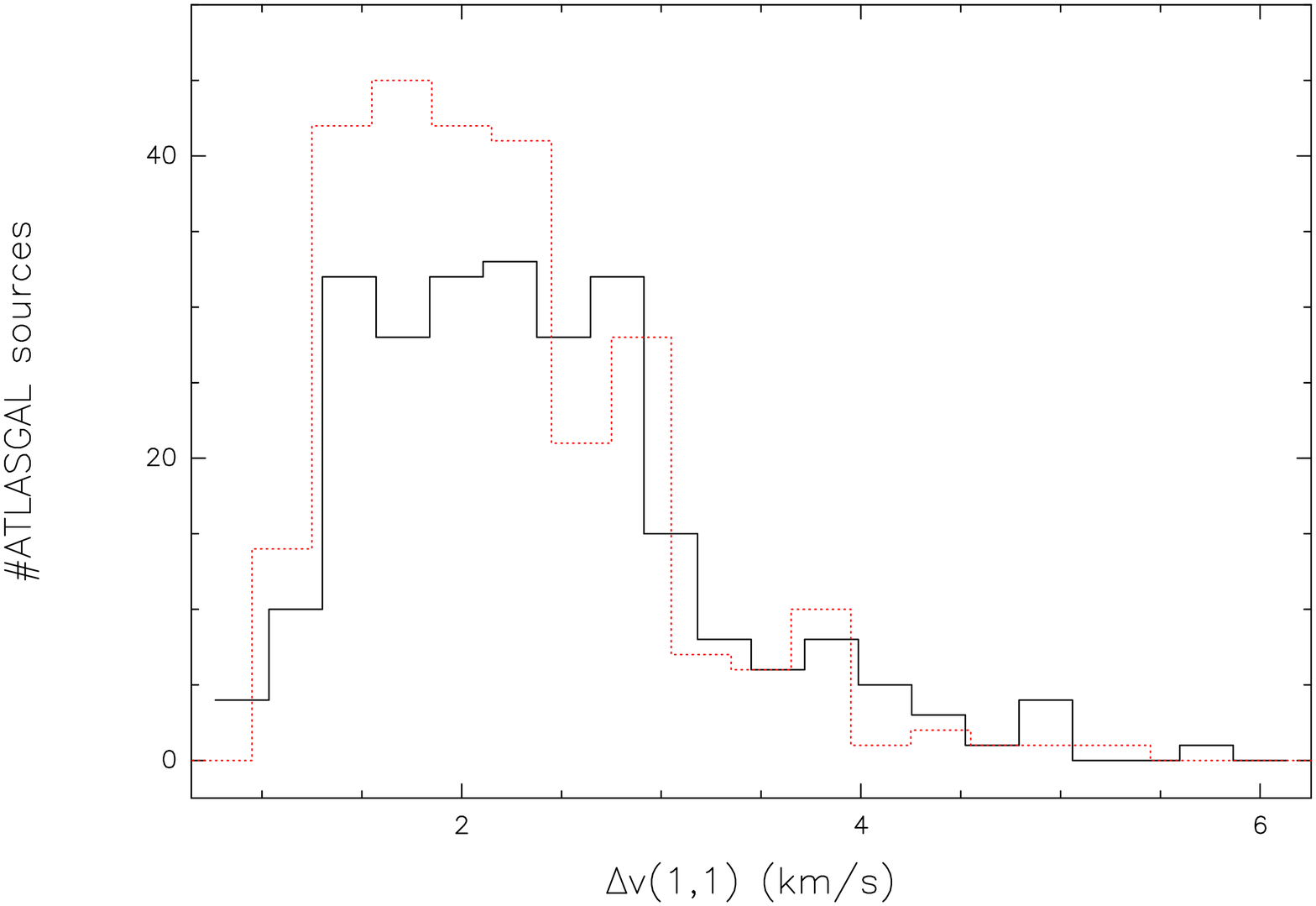}\vspace*{0.5cm}
\includegraphics[angle=0,width=9.0cm]{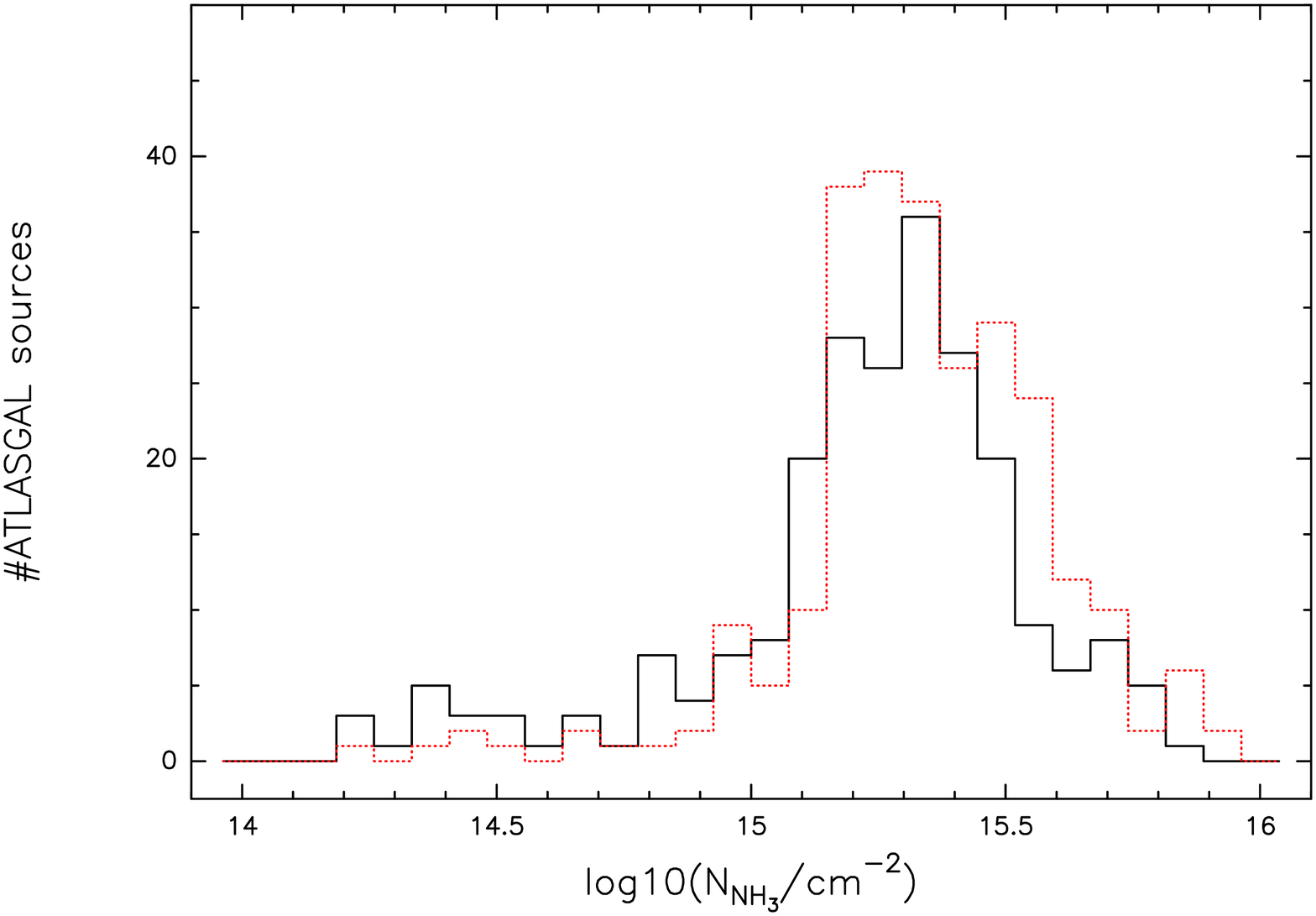}

\includegraphics[angle=0,width=9.0cm]{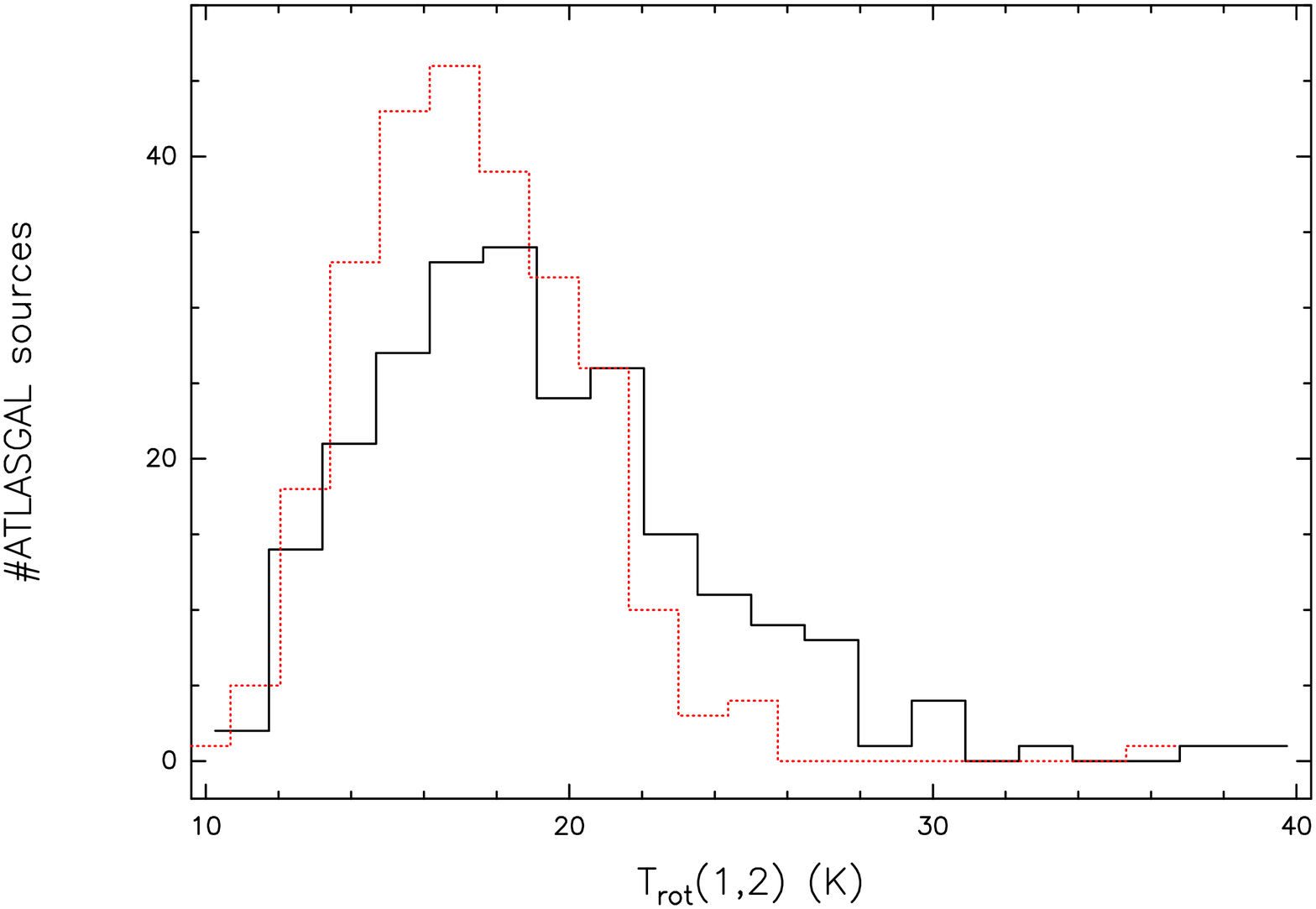}
\caption[Number distribution of sources with NH$_3$ (1,1) linewidth, rotational temperature and column density for NH$_2$D detections and non-detections]{Distribution of ATLASGAL sources with the NH$_3$ (1,1) line width, rotational temperature, and NH$_3$ column density are plotted for NH$_2$D detections in red and non-detections in black.}\label{nh2d-histo-detect}
\end{figure}

\subsection{NH$_2$D detections and non-detections}
We divide the ATLASGAL sources observed in NH$_2$D into two subsamples: one that shows NH$_2$D emission, and another for which NH$_2$D is not detected. Figure \ref{nh2d-histo-detect} shows histograms of the NH$_3$ (1,1) line width, NH$_3$ column density, and rotational temperature for non-detections in black and for detections in red. There is no difference in the line width, column density, or temperature between the two subsamples; they have a peak at $\sim 2.5$ km~s$^{-1}$, $\sim 2 \times 10^{15}$ cm$^{-2}$, and 17 K respectively. The lower panel has a smaller number of sources detected in NH$_2$D than non-detections at rotational temperatures higher than 22 K. Because HII regions have an enhancement in the rotational temperature distribution around 20 K (see \cite{2012A&A...544A.146W}), the lower panel of Fig. \ref{nh2d-histo-detect} suggests that the NH$_2$D detection rate is low in HII regions.

\subsection{NH$_3$ deuteration at different evolutionary phases}
\label{deuteration evolutionary phases}
To distinguish various evolutionary stages of ATLASGAL sources detected in NH$_2$D we followed the classification introduced in \cite{2017A&A...599A.139K} for the TOP100 sample which covers the whole evolutionary sequence. Based on the brightness of these sources at infrared wavelengths and their radio continuum flux, four classes are separated. Adapting this classification, \cite{2018MNRAS.473.1059U} identified the evolutionary stage of the majority of ATLASGAL sources, which we used for association with our NH$_2$D detections. This results in the following: 
\begin{itemize}
\item an ATLASGAL sample of 19 clumps that are \textit{70 $\mu$m weak}: These sources have no pointlike or only a weak counterpart in the Hi-GAL data at 70$\mu$m. This sample is supposed to represent a starless phase or about the earliest stage of high-mass star formation.
\item 32 \textit{MIR-weak} sources: This sample shows compact 70 $\mu$m emission, but is not detected at MIR wavelengths or emits only a weak 24$\mu$m flux below the limit of 2.6 Jy that corresponds to an 8 $M_{\odot}$ star at 4 kpc. 
\item 152 \textit{MIR-bright} clumps: These are identified by their bright compact emission at 8 and 24 $\mu$m. These sources show signs of star formation activity such as infall or outflows.
\item 37 \textit{compact HII regions}: These objects are characterized by a strong MIR and radio continuum flux. They are the latest evolutionary stage, where high-mass protostars emit ultraviolet radiation, and thus heat and ionise their remaining molecular cloud forming compact HII regions.
\end{itemize}
We summarise the identification of the various evolutionary phases of the ATLASGAL sample detected in NH$_2$D in Table \ref{nh2d-samples}. This gives the fraction of sources in each evolutionary stage for the whole sample and shows that most NH$_2$D detections are MIR bright, that the number of MIR-weak sources and compact HII regions are similar, and that the lowest number of NH$_2$D detections are 70 $\mu$m-weak clumps. Comparison of the [NH$_2$D]/[NH$_3$] ratio with the NH$_3$ (1,1) line width and rotational temperature (see Fig. \ref{nh2d-dv11-trot-phases}) shows that the 70 $\mu$m-weak sample exhibits the narrowest line widths ($<$ 2.5 km~s$^{-1}$) and smallest rotational temperatures ($<$ 17.2 K). The MIR-weak and MIR-bright sources cover a large range around the mean line width of 2.2 km~s$^{-1}$ and mean rotational temperature of 17.3 K. Large line widths up to 6.4 km~s$^{-1}$ and high rotational temperatures up to 23 K are found for the compact HII regions. However, the subsamples in the various evolutionary phases do not differ with regard to NH$_3$ deuteration and the [NH$_2$D]/[NH$_3$] ratios do not show any trend with the evolutionary sequence.\\

\begin{figure}[h]
\centering
\includegraphics[angle=0,width=9.0cm]{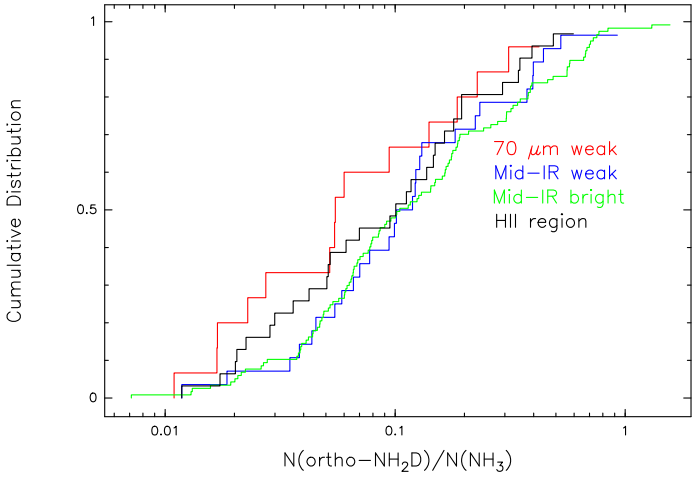}
\caption[Cumulative distribution plot of ortho NH$_2$D/NH$_3$ ratio]{Cumulative distribution functions display [NH$_2$D]/[NH$_3$] ratios for the subsamples in Table \ref{nh2d-samples}. The distribution of 70 $\mu$m-weak sources is marked as a solid red line, the MIR-weak clumps are shown as a dashed blue curve, the MIR-bright phase as a dotted green line, and the compact HII regions as a dashed-dotted black curve. A deuterium fraction of NH$_3$ $> 1$ might result from passing through several dense and cold phases (see Sect. \ref{other samples}). A systematic error of the NH$_3$ deuteration is given by the difference between the NH$_3$ and NH$_2$D excitation temperature.}\label{nh2d-distribution}
\end{figure}

We performed a Kolmogorov-Smirnov (KS) test with the subsamples at the different evolutionary stages to analyse whether or not they differ significantly in NH$_3$ deuteration. The distributions of the 70 $\mu$m-weak sources, the MIR-weak sources, and compact HII regions do not contradict the idea that they are drawn from the same parent population. The cumulative distribution plot in Fig. \ref{nh2d-distribution} yields a higher NH$_3$ deuteration for clumps with weak or bright MIR emission. The NH$_2$D/NH$_3$ ratio is low for the 70 $\mu$m-weak phase, rises for the MIR-weak/MIR-bright sources, and decreases again for the compact HII regions. Compared to low-mass star formation, where deuteration decreases with rising temperature \citep{2005ApJ...619..379C,2009A&A...493...89E}, we find a shift of the maximum deuterated NH$_3$ of ATLASGAL sources to a later evolutionary phase. The initial conditions to form NH$_2$D and to increase the deuterium fractionation is sensitive to (low) temperature and (high) density. While the 70 $\mu$m-weak clump might be cold, it is unclear that it is sufficiently dense. There is evidence from some recent observations that high-mass 70 $\mu$m-weak clumps are likely to contain modest sub-structure in dense cores, and that the clump mass reservoir is dominated by low density material \citep{2019A&A...622A..54P}. During the MIR-weak phase, the density is likely to be high enough while the overall temperatures remain low enough to result in deuterium enhancement.  During the MIR-bright phase, the NH$_2$D/NH$_3$ ratio is likely to reach the peak. Protostellar heating raises the temperature, but affects deuteration only in the immediate vicinity of the core so that the decrease of the deuterium fractionation might be undetectable in single-dish observations. This might be different when zooming into these cores at a resolution of less than a few thousand AU, where the direct heating from the protostar is efficient and would reveal the effect of deuteration \citep{2011A&A...530A.118P}. The continued star formation during the compact HII region phase is expected to result in a decrease in deuteration that makes it undetectable. The exact nature of deuterium enhancement with evolution can only be constrained by high-resolution observations in a single star-forming region that hosts cores at all these various evolutionary stages.

\begin{table*}[htbp]
\caption[]{Number and fraction of ATLASGAL sources detected in NH$_2$D at different evolutionary stages of high-mass star formation.}
\label{nh2d-samples}
\centering
\begin{tabular}{l c c}
\hline\hline
Sample &  Number & Fraction (\%) \\ \hline 
70 $\mu$m-weak  &   19    & 7  \\
MIR-weak   &   32   & 12  \\
MIR-bright  &   152  &  58 \\
Compact HII regions  &  37   & 14  \\ \hline 
\end{tabular}
\end{table*}

\begin{table*}
\begin{minipage}{\textwidth}
\caption[Velocity integrated intensity derived from the NH$_2$D ortho transition, para line and the ratio of both integrated intensities]{Velocity integrated intensity derived from the NH$_2$D line at 86 GHz, from the line at 110 GHz, and the ratio of both integrated intensities. Errors are given in parentheses. The full table is available at CDS.}             
\label{ortho-para}     
\centering                                      
\begin{tabular}{l c c c}         
\hline\hline                       
& $\int T_{\rm mb} \ \rm dv$ (86GHz)  & $\int T_{\rm mb} \ \rm dv$ (110GHz)  & $\int T_{\rm mb} \ \rm dv$ (86GHz)/$\int T_{\rm mb} \ \rm dv$ (110GHz)  \\ 
Name  &  (K km~s$^{-1}$)  & (K km~s$^{-1}$) & (K km~s$^{-1}$)   \\                    
\hline    
%Added by TeX Support                             
G10.21-0.30 & 1.1$(\pm0.1$) & 0.2$(\pm0.1$) & 4.8$(\pm2.4$) \\
G10.67-0.22 & 1.1$(\pm0.1$) & 0.5$(\pm0.1$) & 2.3$(\pm0.6$) \\
G12.89+0.49 & 2.4$(\pm0.2$) & 0.9$(\pm0.1$) & 2.8$(\pm0.4$) \\
G12.50-0.22 & 2.3$(\pm0.1$) & 1.0$(\pm0.1$) & 2.2$(\pm0.3$) \\
G12.90-0.03 & 2.1$(\pm0.2$) & 0.6$(\pm0.2$) & 3.7$(\pm1.0$) \\
G11.92-0.61 & 2.5$(\pm0.2$) & 1.0$(\pm0.2$) & 2.5$(\pm0.4$) \\
G11.94-0.61 & 1.6$(\pm0.2$) & 0.7$(\pm0.2$) & 2.4$(\pm0.6$) \\
G12.91-0.26 & 2.9$(\pm0.1$) & 0.9$(\pm0.1$) & 2.8$(\pm0.4$) \\
G13.12-0.09 & 1.3$(\pm0.2$) & 1.1$(\pm0.2$) & 1.2$(\pm0.3$) \\
G13.90-0.51 &  2.4$(\pm0.1$)  & 0.8$(\pm0.1$) & 3.1$(\pm0.5$) \\
G14.25+0.07 & 1.8$(\pm0.2$) & 0.6$(\pm0.2$) & 3.0$(\pm1.0$) \\
G17.65+0.17 & 2.0$(\pm0.2$) & 0.9$(\pm0.2$) & 2.3$(\pm0.5$) \\
G27.37-0.17 & 4.9$(\pm0.1$) & 1.9$(\pm0.2$) & 2.5$(\pm0.2$) \\
G29.91-0.04 & 1.7$(\pm0.1$) & 1.0$(\pm0.1$) & 1.8$(\pm0.3$) \\

\hline                                            
\end{tabular}
\end{minipage}
\end{table*}

\subsection{The integrated intensity ratio of ortho and para NH$_2$D}
In addition to the calculation of the ortho-to-para ratio given in Sect. \ref{ortho-to-para ratio}, we also determined the velocity integrated intensity ratio, $r$, of the ortho and para NH$_2$D transition to compare with other studies. The integrated intensity of the ortho line, para line, and the ratio of the two is reported in Table \ref{ortho-para}. This resulted in a mean $r$ of $2.6 \pm 0.8$ which confirms the results from previous studies using smaller samples. \cite{2015A&A...575A..87F} obtained a ratio $r$ of $2.6 \pm 0.6$ for high-mass star-forming samples covering different evolutionary phases, which is in agreement with our result. In addition, our values are similar to the integrated line intensity ratios derived by \cite{2007AA...467..207P} for clumps embedded in infrared-dark clouds. \cite{2001ApJ...554..933S} give a range in $N_{\rm tot} (\rm 86GHz) /N_{\rm tot} (\rm 110GHz)$ for low-mass protostellar cores of between approximately $2$ and 6, which is in agreement with our values for $r$. 

\begin{figure}[h]
\centering
\includegraphics[angle=0,width=9.0cm]{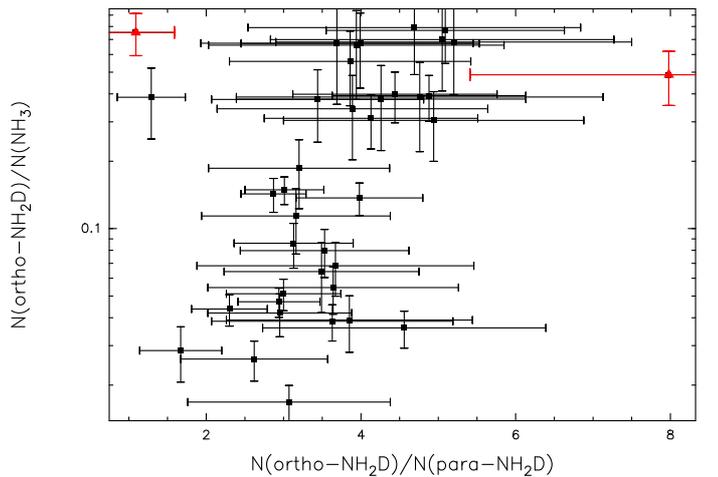}
\caption[Comparison of ortho-to-para ratio and NH$_3$ deuteration]{NH$_3$ deuteration is compared with the ortho-to-para ratio. ATLASGAL sources without hyperfine structure in NH$_2$D at 86 or 110 GHz are illustrated as black points and clumps with detected hyperfine structure in both lines as red triangles.}\label{nh2d-ortho-para}
\end{figure}

\begin{figure}[h]
\centering
\includegraphics[angle=0,width=9.0cm]{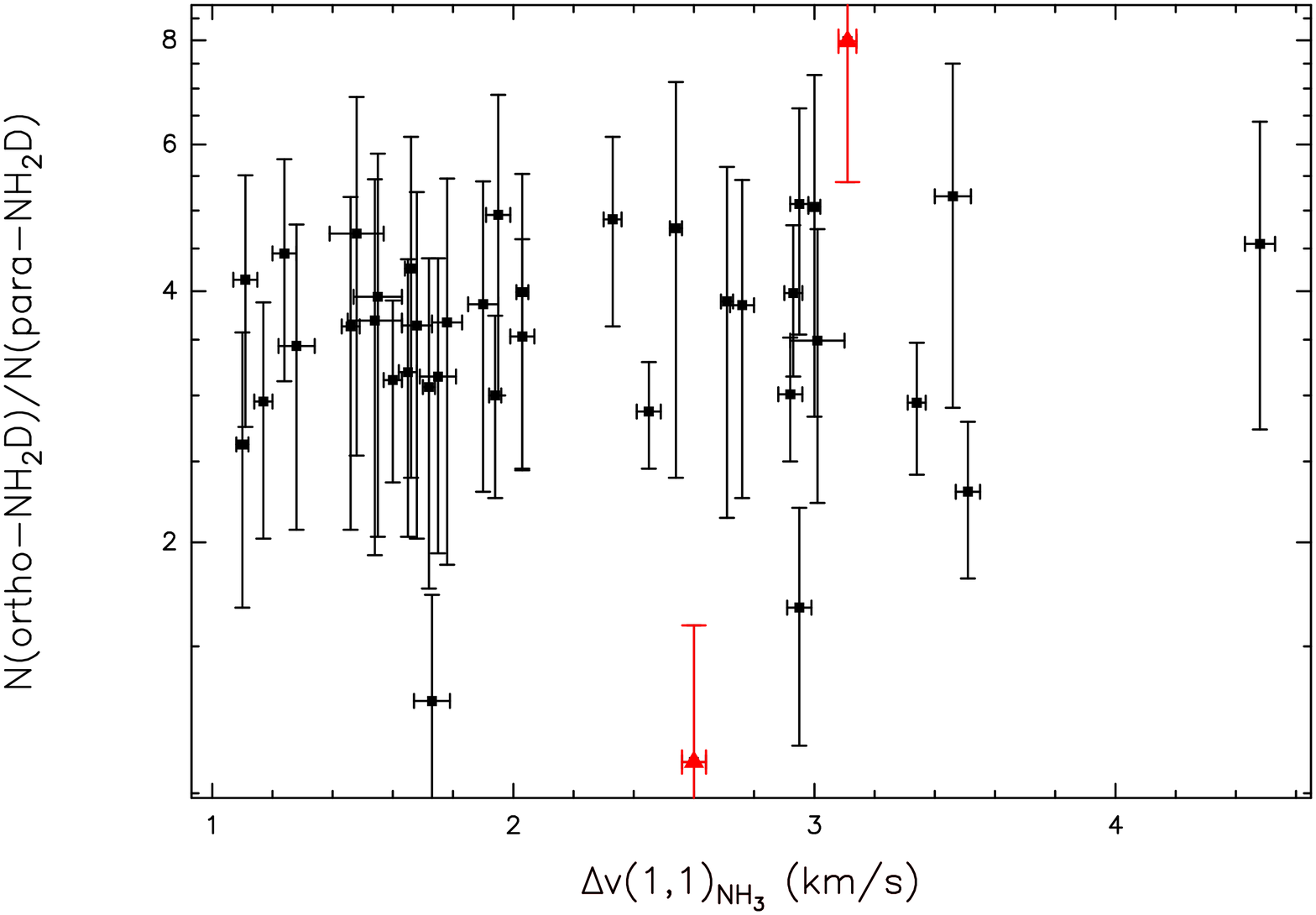}\vspace*{0.5cm}
\includegraphics[angle=0,width=9.0cm]{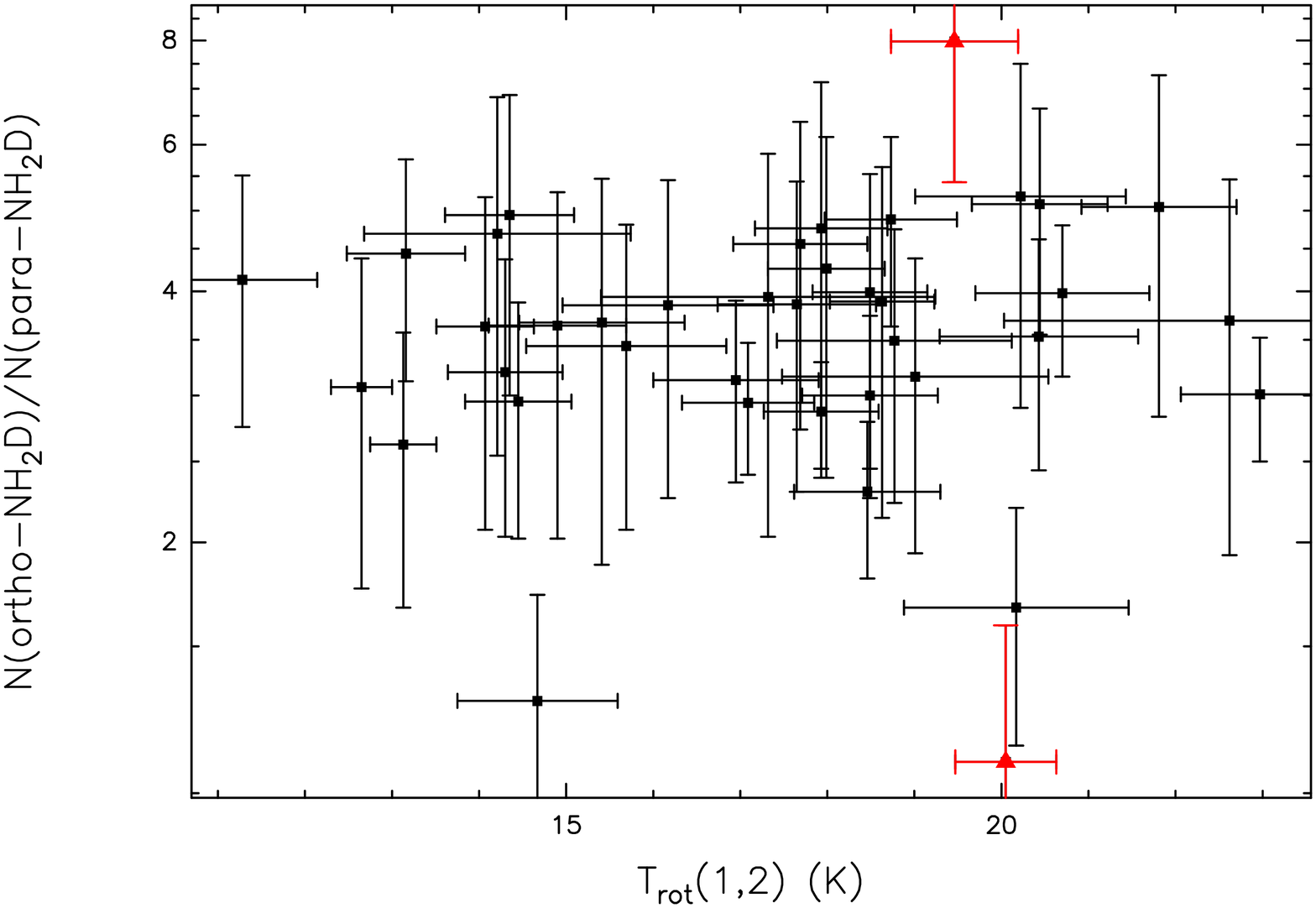}
\caption[Ortho-to-para ratio plotted against NH$_3$ (1,1) linewidth and rotational temperature]{Ortho-to-para ratio against NH$_3$ (1,1) line width (upper panel) and against rotational temperature (lower panel). ATLASGAL clumps, which have no detected hyperfine structure in NH$_2$D at 86 or 110 GHz are displayed as black points and sources with detected hyperfine structure in both transitions as red triangles.}\label{ortho-para-dv11}
\end{figure}

We investigate whether or not there is a correlation of the ortho-to-para ratio computed in Sect. \ref{ortho-to-para ratio} with NH$_3$ deuteration, NH$_3$ (1,1) line width, and rotational temperature in Fig. \ref{nh2d-ortho-para} and \ref{ortho-para-dv11}. ATLASGAL sources, whose hyperfine structure is not detected in ortho or para NH$_2$D, are indicated as black points. There are only two clumps with detected hyperfine structure in NH$_2$D at 86 and 110 GHz, which are shown as red triangles. However, Fig. \ref{nh2d-ortho-para} and \ref{ortho-para-dv11} do not show a correlation between the ortho-to-para ratio
and any of the NH$_3$ parameters. This yields therefore no evolutionary trend of  the ortho-to-para ratio.

\section{Conclusions}
\label{conclusions}
Using the Mopra telescope and the IRAM 30m telescope, the 86 GHz NH$_2$D lines were observed toward 992 dust condensations identified in the ATLASGAL survey. In the first quadrant of the Galaxy, 373 sources were also observed in the 110 GHz para line with the IRAM 30m telescope. They are located within a Galactic longitude from 8$^{\circ}$ to 60$^{\circ}$ and between 300$^{\circ}$ and 359$^{\circ}$ and latitude of $\pm 1.5^{\circ}$. We summarise our main results in this section.
\begin{enumerate}
 \item The detection rate of NH$_2$D towards the ATLASGAL sample is high and therefore yields a large NH$_3$ deuteration of these sources.
 \item We calculate the total NH$_2$D column density to determine the NH$_2$D-to-NH$_3$ column density ratio. This results in a large range of NH$_3$ deuteration from 0.007 to 1.6. The deuterium fraction of NH$_3$ in ATLASGAL clumps is higher than the [NH$_2$D]/[NH$_3$] ratios derived for low-mass star-forming samples and in agreement with those obtained in other high-mass star-forming regions. We measure the highest NH$_3$ deuteration reported in the literature so far.
 \item The excitation of NH$_2$D was studied using the transitions at 74 GHz and 110 GHz for the first time to our knowledge. This shows a clear difference between the NH$_2$D and NH$_3$ rotational temperatures for a subsample of the sources. In cases where NH$_2$D temperatures are lower than NH$_3$ temperatures, deuteration would be overestimated, suggesting non-LTE conditions. To determine the NH$_2$D temperature directly, the NH$_2$D lines at 74 and 110 GHz should be observed simultaneously.
 \item Comparison of NH$_2$D detections and non-detections suggests that the fraction of sources detected in NH$_2$D is higher for the earlier evolutionary phases.
 \item We analyse whether or not there is a trend of NH$_3$ deuteration with evolutionary tracers. While the [NH$_2$D]/[NH$_3$] ratio is expected to decrease with rising rotational temperature \citep{2000A&A...361..388R,2015A&A...578A..55S}, we obtain a flat distribution of the deuterium fractionation with the NH$_3$ (1,1) line width, rotational temperature, and the MSX 21 $\mu$m flux. Observations of ATLASGAL clumps with a high [NH$_2$D]/[NH$_3$] ratio within a large beam width might also include cores of large line widths and temperatures. An average over the beam width would lead to an increase of these NH$_3$ properties with an enhanced NH$_3$ deuteration. Future interferometric follow-up observations could resolve this issue.
 \item We divide the ATLASGAL sample into different evolutionary phases, but do not find any correlation between these and NH$_3$ deuteration. The NH$_2$D/NH$_3$ ratio is maximum during the MIR bright phase and is therefore reached at a later evolutionary stage compared to low-mass star formation.
 \item We estimate the ratio of the ortho-to-para NH$_2$D column density ratio. 
 This results in a median ortho-to-para ratio of 3.7 close to the expected value of 3. The ortho-to-para column density ratios are in agreement with those of other low- and high-mass star-forming samples.
\end{enumerate}

\noindent
\textit{Acknowledgements.} M. Wienen acknowledges funding from the European Union’s Horizon 2020 research and innovation programme under the Marie Sk{\l}odowska-Curie grant agreement No 796461.

This paper is dedicated to the memory of Malcolm Walmsley, who passed away before this study could be completed. The present work benefited greatly from his insight and extensive advice. We are grateful for numerous inspiring discussions with him about various aspects of ammonia and its deuteration.

\bibliography{paperslibrary}
\bibliographystyle{aa}

\begin{appendix}
\section{NH$_3$ deuteration and evolutionary tracers}
\begin{figure}[h]
\centering
\includegraphics[angle=0,width=9.0cm]{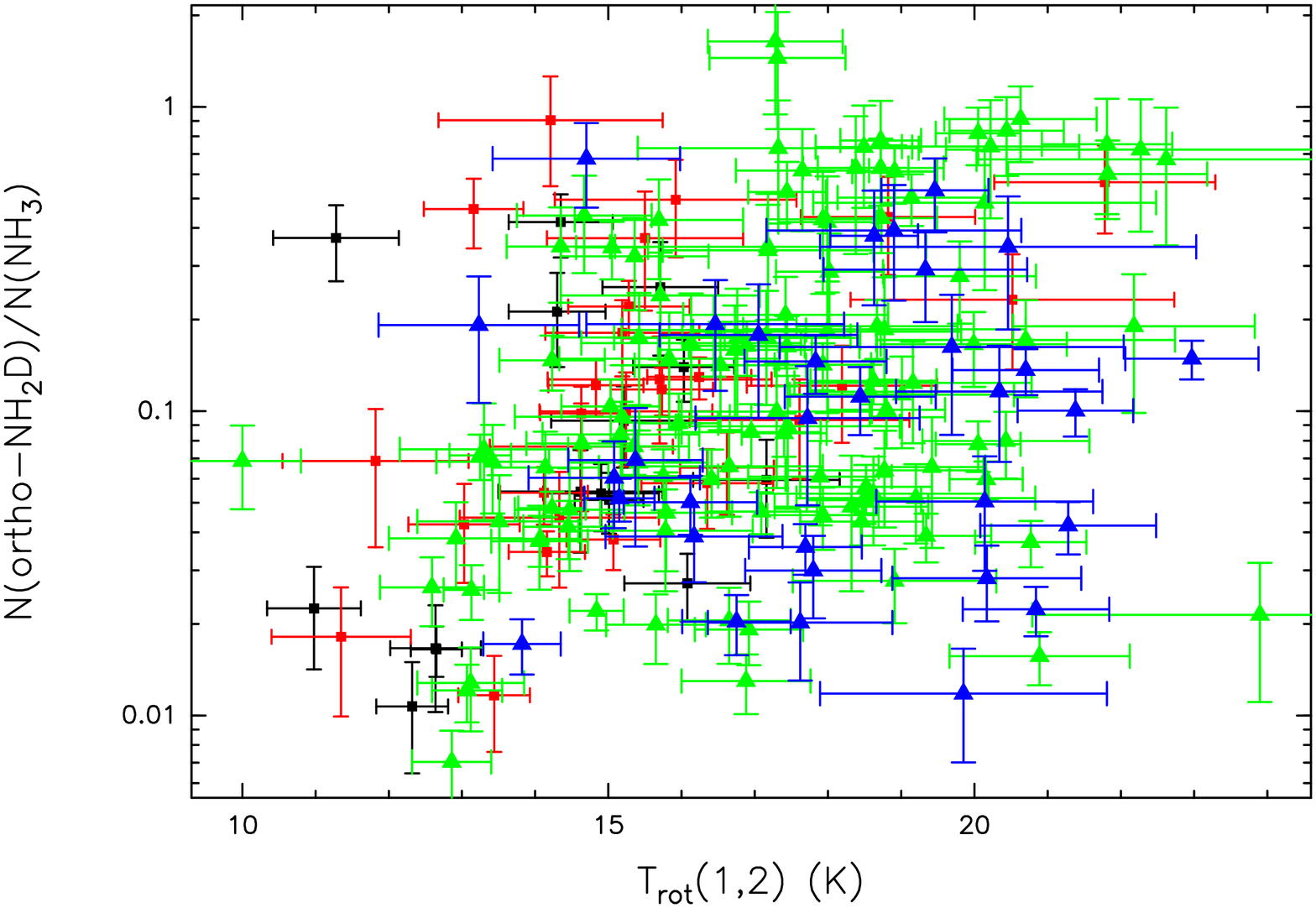}\vspace*{0.5cm}
\includegraphics[angle=0,width=9.0cm]{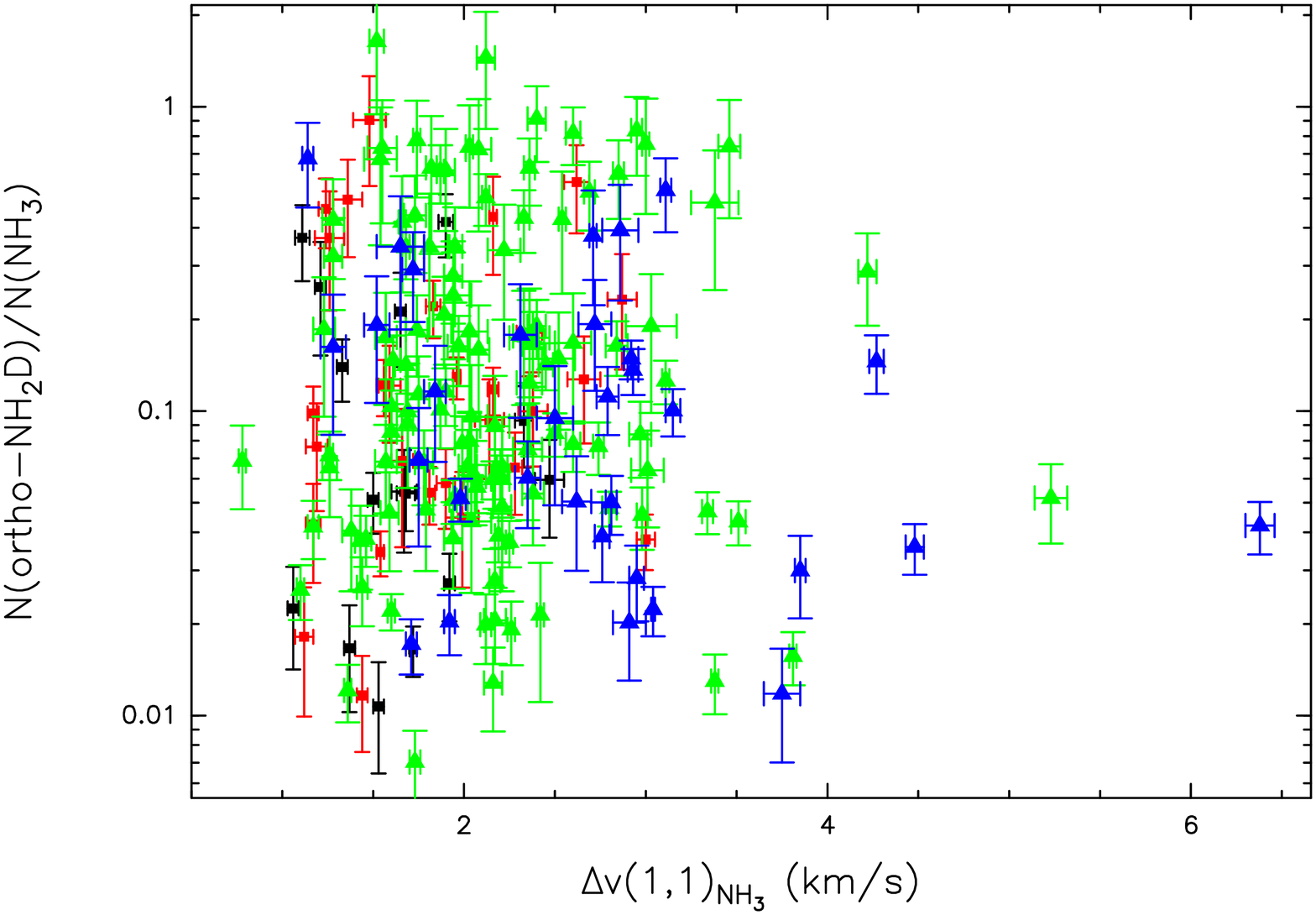}
\caption[NH$_3$ deuteration plotted against NH$_3$ (1,1) linewidth and rotational temperature]{NH$_3$ deuteration as a function of NH$_3$ (1,1) line width and rotational temperature between the (1,1) and (2,2) inversion transition. ATLASGAL clumps that are 70 $\mu$m weak are illustrated as black points, MIR-weak sources as red points, MIR-bright clumps as green triangles, and compact HII regions as blue triangles.}\label{nh2d-dv11-trot-phases}
\end{figure}

We associated our NH$_2$D detections with the ATLASGAL sources classified in \cite{2018MNRAS.473.1059U}. The [NH$_2$D]/[NH$_3$] ratio of the different subsamples is plotted against the NH$_3$ rotational temperature and (1,1) line width in Fig. \ref{nh2d-dv11-trot-phases}.

\section{Do NH$_2$D clumps with and without detected hyperfine structure differ?}
\begin{figure}
        \centering
        \includegraphics[angle=0,width=9.0cm]{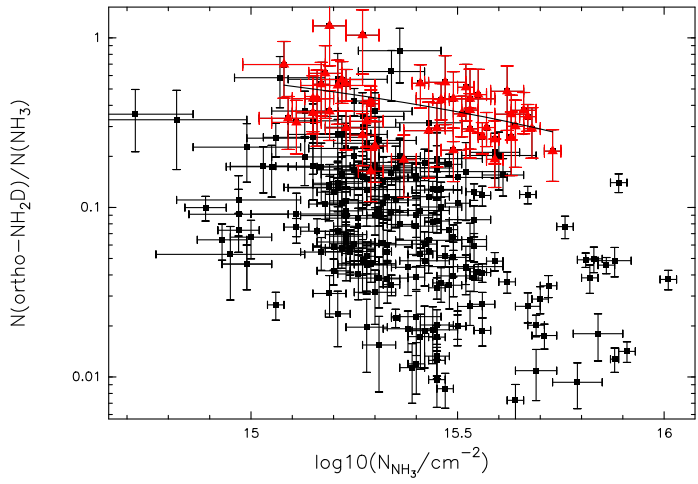}
        \caption[Dependence of NH$_3$ deuteration on logarithm of NH$_3$ column density]{Dependence of NH$_3$ deuteration on NH$_3$ column density. A linear fit is illustrated by the straight line. ATLASGAL clumps without detected hyperfine structure in NH$_2$D are illustrated as black points and %optically thick sources
        sources with detected hyperfine structure as red triangles.}\label{nh2d-nh3-col}
\end{figure}

After not seeing any trend in the NH$_3$ deuteration with the evolution of our ATLASGAL subsample, we searched for differences of ATLASGAL sources with or without detected hyperfine structure in NH$_2$D. We cannot distinguish these two categories based on the rotational temperature, NH$_3$ line width, or MSX 21$\mu$m flux of the ATLASGAL clumps as indicated by Fig. \ref{nh2d-dv11-trot}. We compared the [NH$_2$D]/[NH$_3$] ratio with the NH$_3$ column density in Fig. \ref{nh2d-nh3-col} to examine whether or not the largest column densities are related to the high deuteration of clumps with detected hyperfine structure and the sources without detected hyperfine structure and with low deuteration exhibit the lowest column densities. However, we find no difference in the column density of ATLASGAL clumps with or without detected hyperfine structure. Because Fig. \ref{nh2d-nh3-col} might present a decreasing trend in the deuterium fraction of NH$_3$ of sources with detected hyperfine structure and rising NH$_3$ column density, we performed a t-test to examine whether or not the slope of the distribution is equal to zero. Because the t-test yields a p-value of 0.02, which is below the assumed significance level of 0.05, we can reject the hypothesis that the distribution can be fitted by a function with a slope of zero. The fractionation ratio of the sources with detected hyperfine structure will therefore become lower if the NH$_3$ column density increases and can be fitted by [NH$_2$D]/[NH$_3$] = -0.39 log$_{10}$(N$_{\rm tot}$ (NH$_3$)) + 6.5.

\end{appendix}

\end{document}